\documentclass{emulateapj}
\usepackage{verbatim,graphicx,natbib}
\shortauthors{Cowan et al.}
\shorttitle{Thermal Phase Variations of WASP-12b}
\begin{document}

\title{Thermal Phase Variations of WASP-12b: Defying Predictions}
\author{Nicolas B. Cowan\altaffilmark{1,2}, Pavel Machalek\altaffilmark{3,4}, Bryce Croll\altaffilmark{5,6}, Louis M. Shekhtman\altaffilmark{1}, Adam Burrows\altaffilmark{7}, Drake Deming\altaffilmark{8,9}, Tom Greene\altaffilmark{4}, Joseph L. Hora\altaffilmark{10}}

\altaffiltext{1}{Center for Interdisciplinary Exploration and Research in Astrophysics (CIERA) and Department of Physics \& Astronomy, Northwestern University, 2131 Tech Dr, Evanston, IL, 60208, USA\\
email: n-cowan@northwestern.edu}
\altaffiltext{2}{CIERA Postdoctoral Fellow}
\altaffiltext{3}{SETI Institute, 189 Bernardo Ave., Suite 100, Mountain View, CA 94043, USA}
\altaffiltext{4}{NASA Ames Research Center, Moffett Field, CA 94035, USA}
\altaffiltext{5}{Department of Astronomy \& Astrophysics, University of Toronto, 50 George St., Toronto, ON, M5S~3H4, Canada}
\altaffiltext{6}{Department of Physics, Massachusetts Institute of Technology, Cambridge, MA, 02139, USA}
\altaffiltext{7}{Department of Astrophysical Sciences, Princeton University, Princeton, NJ 05844, USA}
\altaffiltext{8}{Planetary Systems Laboratory, NASA Goddard Space Flight Center, Greenbelt, MD, 20771, USA}
\altaffiltext{9}{Department of Astronomy, University of Maryland, CSS Bldg., Rm. 1204, Stadium Dr., College Park, MD, 20742, USA}
\altaffiltext{10}{Harvard-Smithsonian Center for Astrophysics, Cambridge, MA 02138, USA}

\begin{abstract}
We report \emph{Warm Spitzer} full-orbit phase observations of WASP-12b at 3.6 and 4.5 $\mu$m. This extremely inflated hot Jupiter is thought to be overflowing its Roche lobe, undergoing mass loss, accretion onto its host star, and has been claimed to have a C/O ratio in excess of unity.   We are able to measure the transit depths, eclipse depths, thermal and ellipsoidal phase variations at both wavelengths.   
The large amplitude phase variations, combined with the planet's previously-measured day-side spectral energy distribution, is indicative of non-zero Bond albedo and very poor day--night heat redistribution. 
The transit depths in the mid-infrared ---$(R_{p}/R_{*})^{2} = 0.0123(3)$ and $0.0111(3)$ at 3.6 and 4.5 $\mu$m, respectively--- indicate that the atmospheric opacity is greater at 3.6 than at 4.5~$\mu$m, in disagreement with model predictions, irrespective of C/O ratio. The secondary eclipse depths are consistent with previous studies:
$F_{\rm day}/F_{*} = 0.0038(4)$ and 0.0039(3) at 3.6 and 4.5 $\mu$m, respectively. We do not detect ellipsoidal variations at 3.6 $\mu$m, but our parameter uncertainties ---estimated via prayer-bead Monte Carlo--- keep this non-detection consistent with model predictions. At 4.5 $\mu$m, on the other hand, we detect ellipsoidal variations that are much stronger than predicted. If interpreted as a geometric effect due to the planet's elongated shape, these variations imply a 3:2 ratio for the planet's longest:shortest axes and a relatively bright day--night terminator. 
If we instead presume that the 4.5 $\mu$m ellipsoidal variations are due to uncorrected systematic noise and we fix the amplitude of the variations to zero, the best fit 4.5 $\mu$m transit depth becomes commensurate with the 3.6 $\mu$m depth, within the uncertainties. The relative transit depths are then consistent with a Solar composition and short scale height at the terminator.   Assuming zero ellipsoidal variations also yields a much deeper 4.5 $\mu$m eclipse depth, consistent with a Solar composition and modest temperature inversion. We suggest future observations that could distinguish between these two scenarios.     
\end{abstract}

\section{Introduction}
Thermal phase variations are a powerful way to constrain the climate on exoplanets. Such observations have been made for non-transiting short-period planets \citep{Cowan_2007, Crossfield_2010}, but are most potent when combined with transit and eclipse observations for edge-on systems, because of the additional knowledge of the planet's inclination, mass and radius \citep{Knutson_2007a, Knutson_2009a, Knutson_2009b}. Secondary eclipse depths provide a constraint on the planet's day-side temperature. Thermal phase variations probe the day--night temperature contrast and hence the planet's heat redistribution efficiency. If the observational cadence and signal-to-noise ratio are sufficiently high, phase variations are also sensitive to the offset between the noon meridian and the planet's hottest local stellar time, hence constraining wind speed and direction. 

By considering eclipse depths at a variety of wavelengths for a sample of 24 transiting planets, \cite{Cowan_2011b} estimated their day-side effective temperatures, hence placing a joint constraint on the Bond albedo and heat recirculation efficiency of these planets. That study found that typical hot Jupiters exhibit a variety of albedo/recirculation efficiencies, but planets with substellar equilibrium temperatures greater than $T_{0}\approx2700$~K all seem to have lower albedo and/or recirculation efficiency. In other words, the hottest transiting giant planets have a qualitatively different climate than the merely hot Jupiters, but it is not known whether this is due to a difference in albedo, circulation, or both. Direct measurements of hot Jupiter geometric albedos from optical secondary eclipse observations span more than an order of magnitude and do not resolve this degeneracy.

In this paper, we break the albedo-recirculation degeneracy for WASP-12b \citep{Hebb_2009}, one of the very hottest known exoplanets, with a day-side temperature of $\sim3000$~K: the amplitude of thermal phase variations is a direct measure of the planet's day-night temperature contrast and hence heat transport efficiency.  If the night-side temperature is high, then the planet's albedo must be exceedingly low to be consistent with its high day-side temperature.  If, on the other hand, the night-side temperature is low, then the planet has an albedo in the tens of percent. 

WASP-12b has been a fascinating planet since its discovery. The discrepant timing of its secondary eclipse indicated that the planet had a slight eccentricity \citep{Lopez-Morales_2010}, but subsequent eclipse \citep{Campo_2011, Croll_2011} and radial velocity \citep{Husnoo_2011} observations have all but ruled this out. Nevertheless, the planet's short-period orbit (1.1 days; just outside its star's Roche limit) and inflated radius (1.8 $R_{J}$) led to the prediction that it is tidally distorted \citep{Ragozzine_2009, Leconte_2011, Budaj_2011}, and undergoing Roche-lobe overflow followed by accretion onto its host star \citep{Li_2010, Lai_2010}. The putative early ingress of an ultraviolet transit observed by \cite{Fossati_2010} seems to support this prediction, but may also be explained in terms of a leading bow shock from material streaming off the planet \citep{Vidotto_2010, Llama_2011}. 

More recently, \cite{Madhusudhan_2011} used the wavelength-dependance of mid-infrared eclipse depths of \cite{Croll_2011} and \cite{Campo_2011} to constrain the atmospheric composition of WASP-12b, and found it has a carbon to oxygen ratio (C/O) greater than unity, unlike Solar System planets, or the assumed composition of extrasolar planets. Those findings rested heavily on the relative eclipse depths at 3.6 and 4.5 $\mu$m. Our observations of eclipses and transits at these wavebands should be able to reinforce or rule out the high C/O scenario.

\section{Observations \& Reduction}
We acquired observations of WASP-12 (spectral type F9V) with IRAC \citep{Fazio_2004} on the Spitzer Space Telescope \citep{Werner_2004} at 3.6 $\mu$m (2010 Nov 17--18) and 4.5 $\mu$m (2010 Dec 11--12) as part of the Warm Mission. We used the sub-array mode and acquired images every 2~s (1.92~s effective exposure time), observing the system for approximately 1.3 days at each waveband, from slightly before a secondary eclipse to shortly after the following secondary eclipse. This yields 902 data cubes (64 frames of $32\times32$ pixels) at each waveband. 

Due to a scheduling error, we did not observe all of the second eclipse's egress at 3.6 $\mu$m. This does not severely affect our science objectives since we simultaneously fit both eclipses at a given wavelength; even at 3.6 $\mu$m we have nearly two full eclipse lightcurves to work with. 

We use the BCD files and convert MJy/str to electron counts by multiplying the flux values by \emph{GAIN}$\times$\emph{EXPTIME}/\emph{FLUXCONV}, using parameter values from the header of the fits files. We use the \emph{BMJD\_OBS} and \emph{FRAMTIME} parameter values to compute the BJD time at the middle of every exposure.

We start by considering the pixel-by-pixel time-series for each 64-frame data cube, replacing \emph{NaN}'s with the pixel's median over that data cube; if the entire time-series for a given pixel is bad, it is flagged as a bad pixel and ignored in the subsequent analysis.  At 4.5 $\mu$m, the first row of pixels ($y=0$) is consistently bad.  

\cite{Deming_2011} noted that \emph{Warm Spitzer} sub-array data cubes exhibit a frame-dependent background flux.  At 3.6 $\mu$m, the first and 58th frames are consistent background outliers (both high), and there is a clear drop in background flux throughout each data cube \cite[see Figure~5 of][]{Deming_2011}.  At 4.5 $\mu$m, the same two frames (1 and 58) are outliers (high and low backgrounds, respectively), and there is a slight \emph{increase} in background flux throughout each data cube.  We elect to ignore the 1$^{\rm st}$ and 58$^{\rm th}$ frames of each data cube (3\% of our data). To correct for the gradual change in background flux, we perform initial background subtraction on each frame of the data cube using the IDL routine \verb+MMM+ and excluding the 16 central pixels of the detector (those closest to the star). 

We then perform a two-step sigma clipping on each pixel's time-series, replacing $4\sigma$ outliers by the pixel's median in that data cube.  The sigma-clipping at the pixel level affects 0.028\% and 0.035\% of our science time-series data at 3.6 and 4.5 $\mu$m, respectively.

To determine the centroid of our target, we first identify the brightest of the central 16 pixels in each frame, then fit a 2D Gaussian to the 7x7 pixel box centered on this brightest pixel using the IDL routine \verb+GAUSS2DFIT+.\footnote{We follow \cite{Agol_2010}, who compared many centroiding algorithms and found this one to be optimal. Using flux-weighted centroiding instead of PSF-fitting results in slightly worse $\chi^{2}$, commensurate correlated noise as measured using $\beta$ (see first Section~4.1), and consistent astrophysical parameters.} 

We perform aperture photometry on the individual frames of the data cubes using the IDL routine \verb+APER+ with a sky annulus of 7--12 pixels in radius \citep[as did][who observed the same system with the same instrument]{Campo_2011}. Since we perform an initial background subtraction early in our reduction pipeline (see above), our results are not very sensitive to changes in the sky annulus.  We verified that nudging the inner/outer radius of the annulus by 1--2 pixels does not significantly affect the goodness-of-fit or astrophysical parameters. 

Bad pixels are ignored in the background estimation; images with a bad pixel within the aperture are ignored. To determine the optimal aperture for our analysis, we re-run our entire data reduction and analysis pipeline for a range of apertures, from 1.5 to 5.0 pixels, in increments of 0.5 pixel.  We find that for both wavebands, the root-mean-squared scatter in the residuals are minimized for an aperture of 2.5 pixels, which we therefore adopt for the remainder of our study. This is smaller than the apertures of 3.75 and 4.0 pixels (for 3.6 and 4.5 $\mu$m images, respectively) used by \cite{Campo_2011}. While using a larger aperture might reduce the photon-counting uncertainty, a small aperture makes it easier to correct for our dominant source of systematic uncertainty, the intra-pixel sensitivity variations described in Section 3.2.  Since the \emph{Spitzer} heater cycling was different for the two observing campaigns (see Section~3.2), it is possible that the nature of this systematic was different for the \cite{Campo_2011} observations.

Lastly, we perform an iterative 4$\sigma$ clipping on the flux time-series, removing any outliers.  This affects 0.01\% and 0.02\% of the science time-series data at 3.6 and 4.5 $\mu$m, respectively. The 3.6 $\mu$m cut is more generous because of the greater systematic flux variations, as described below.

\begin{figure*}[htb]
\includegraphics[width=170mm]{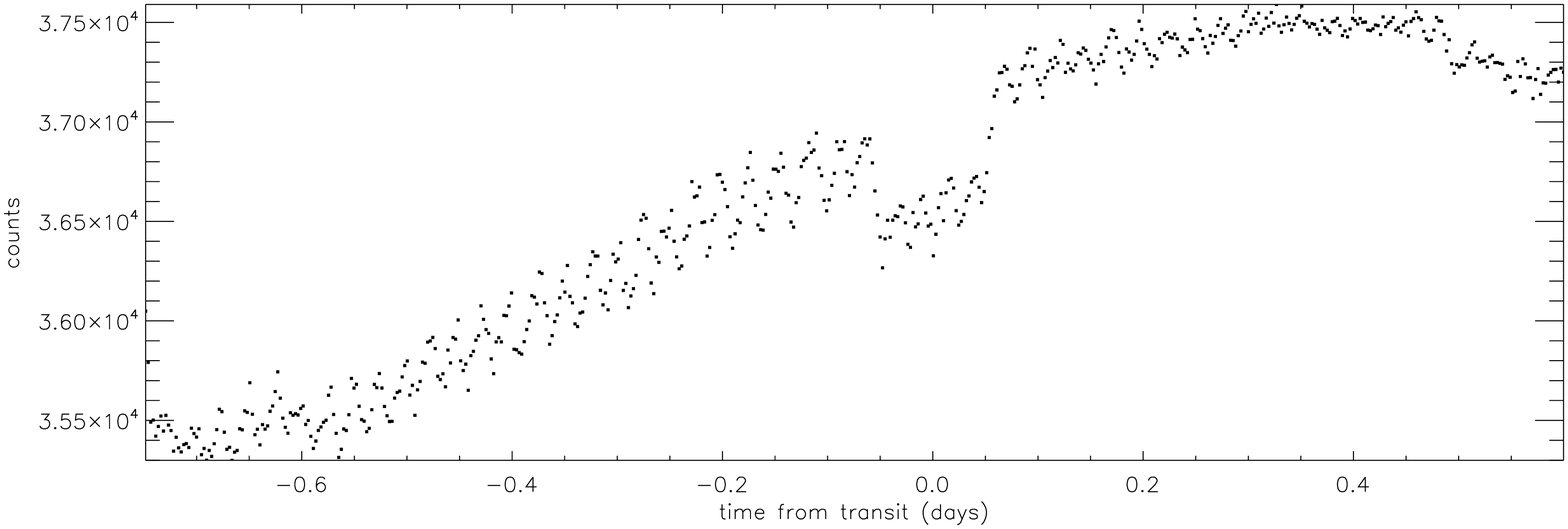}\\
\includegraphics[width=170mm]{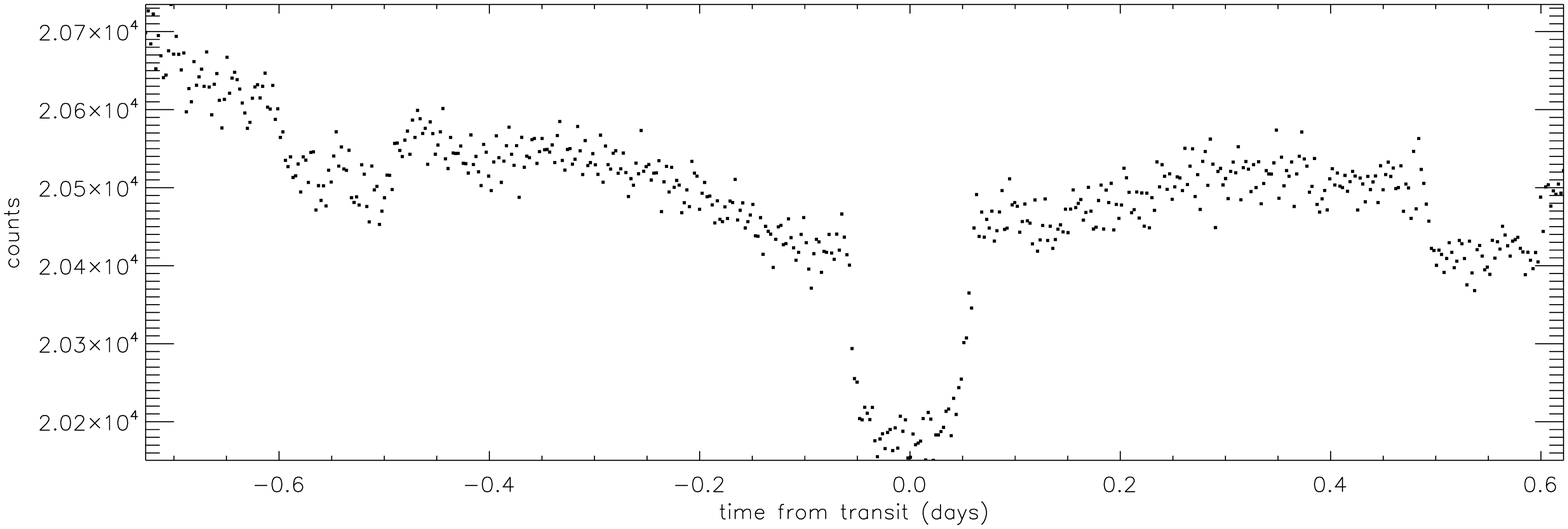}
\caption{Raw photometry in electron counts at 3.6 $\mu$m (top) and 4.5 $\mu$m (bottom). The data have been binned for ease of viewing.}
\label{wasp_12_raw}
\end{figure*}

The raw photometry is shown in Figure~\ref{wasp_12_raw}.  (Note that we perform all of our analysis on the unbinned data, but we bin the data for plotting.)   The transits are easy to spot in the middle of the observations.  The eclipses and phase variations are also visible by eye at 4.5 $\mu$m.  For the 3.6 $\mu$m light curve, the first eclipse and the phase variations are difficult to distinguish from detector systematics by eye. We estimate the system flux in mJy by converting back to MJy/str and using the pixel scale parameter values, \emph{PXSCAL1} and \emph{PXSCAL2}. Our system fluxes ---23.0(5)~mJy and 14.7(1)~mJy, at 3.6 and 4.5 $\mu$m, respectively--- are approximately 10\% lower than those of \cite{Campo_2011}, even when we adopt their larger apertures.  \cite{Fazio_2004} expected the absolute photometric precision of IRAC to be better than 10\%.

As part of our \emph{Warm Spitzer} observations, we also obtained mapping data in sub-array mode for WASP-12, at each 3.6 and 4.5 $\mu$m, immediately following the time series described above. For the mapping data we acquired images every 0.4~s (0.36~s effective exposure time); we obtained 450 data cubes in each waveband. The purpose of these data was to map the central four pixels of the detector by scanning over them in 0.2-pixel and 0.1-pixel increments in the $x$- and $y$-directions, respectively. 

The data reduction for the mapping data is identical to that for the science lightcurve, except that we stack the frames of each data cube using a pixel-by-pixel median. Aperture photometry is performed on these 450 stacked images rather than on the individual frames.  This is necessary because of the shorter integration times for these data. (We also tried this data reduction ---performing photometry cube-by-cube rather than frame-by-frame--- on our phase curve data.  Our best-fit models parameters were consistent using this approach, but the detector systematics were harder to correct for, leading to larger parameter uncertainties.)

The pixel-level sigma-clipping affects 0.01\% of both the 3.6 and 4.5 $\mu$m mapping data.  There are no outliers in the flux time series for the mapping data.

\section{Model}
Our model has 9 free astrophysical parameters, plus up to 11 free parameters to characterize the detector response. The model parameters are listed in Table~\ref{variables} and described below.
\begin{deluxetable}{lr}
\tabletypesize{\scriptsize}
\tablecaption{Model Variables \label{variables}}
\tablewidth{0pt}
\tablehead{
\colhead{Name} & \colhead{Symbol}}
\startdata
Stellar Flux & $F_{*}$\\
Orbital Period$^{a}$ & $P$\\
Impact Parameter & $b$\\
Geometric Factor& $a/R_{*}$ \\
Time of Transit (BMJD) & $t_{0}$ \\
Area Ratio & $(R_{p}/R_{*})^{2}$ \\
Mean Planet Flux & $\langle F_{p}/F_{*}\rangle$ \\
Thermal Phase Amplitude & $A_{\rm therm}$ \\
Phase Offset & $\alpha_{\rm max}$  \\
Ellipsoidal Amplitude & $A_{\rm ellips}$ \\
t linear$^{b}$ & $m_{t}$ \\
$x$ Linear$^{c}$ & $a_{1}$ \\
$y$ Linear$^{c}$ & $b_{1}$  \\
$x$ Quadratic$^{c}$ & $a_{2}$  \\
$y$ Quadratic$^{c}$ & $b_{2}$  \\
$x$ Cubic$^{c}$ & $a_{3}$  \\
$y$ Cubic$^{c}$ & $b_{3}$  \\
$x$ Quartic$^{c}$ & $a_{4}$  \\
$y$ Quartic$^{c}$ & $b_{4}$  \\
$x$ Quintic$^{c}$ & $a_{5}$  \\
$y$ Quintic$^{c}$ & $b_{5}$ 
\enddata
\tablenotetext{a}{We fix the orbital period to the value from \cite{Maciejewski_2011}.}
\tablenotetext{b}{This parameter is only used when fitting occultations independently of the rest of the lightcurves.}
\tablenotetext{c}{These parameters are only used in the polynomial IPSV fits described in Section~\ref{simul}.}
\end{deluxetable}

\subsection{Astrophysical Model}
\emph{Occultations:} Transits are modeled using the IDL implementation of \cite{Mandel_2002} with fixed non-linear limb-darkening.  To determine the limb-darkening coefficients, we fit a four-parameter \cite{Claret_2000} model to a Kurucz stellar model with $[Fe/H] = 0.3$, $T_{\rm eff}=6250$~K, and $\log g=4.5$ \citep{Kurucz_1979, Kurucz_2005}.
We set the eccentricity to zero, and fix the orbital period to the value from \cite{Maciejewski_2011}.  The impact parameter, $b$, and the geometrical factor, $a/R_{*}$, are allowed to vary freely.
Eclipses are modeled using the uniform-disk version of the \cite{Mandel_2002} expressions, and both eclipses are modeled simultaneously with the same parameters. This means that we do not look for eclipse depth variability \citep[searches for such variability have so far only resulted in upper limits:][]{Agol_2010, Knutson_2011}. We account for light travel time within the system, but this is only a matter of 22.5~s and does not affect our analysis.  By the same token, we neglect eclipse mapping effects for the planet \citep{Williams_2006, Rauscher_2007, Agol_2010}, since we are insensitive to the resulting offset in eclipse time of less than a minute, let alone the detailed ingress/egress morphology.

\emph{Diurnal Phases:} The planet's  temperature and hence brightness vary as a function of local stellar time. This inhomogenous intensity is modeled with three parameters: 
the orbit-averaged planet/star flux ratio, $\langle F_{p}/F_{*}\rangle$, the semi-amplitude of thermal phase variations, $A_{\rm therm}$, and the offset of the phase peak from superior conjunction, $\alpha_{\rm max}$ ($\alpha_{\rm max}<0$ corresponds to a peak prior to superior conjunction and therefore to an eastward-advected hotspot and super-rotating winds). The phase variations have a sinusoidal shape, corresponding to a sinusoidal longitudinal brightness profile for the planet \citep{Cowan_2008}.\footnote{In general, the offset between thermal phase maximum and superior conjunction is \emph{not} the same as the offset of the hottest longitude of the planet with respect to the sub-stellar meridian. For realistic longitudinal temperature profiles, the observed offset in the lightcurve is significantly greater than the hot spot offset \citep{Cowan_2011a}.  In the case of a first-order sinusoidal phase curve, however, the two offsets are one and the same.} Note that in the limit of poor recirculation, the longitudinal temperature profile should be more akin to a half-sine (uniformly dark on the night-side), leading to thermal phase variations similar to the Lambert phase function: broader minimum, briefer maximum. We neglect reflected star light, which does not contribute appreciably at these wavelengths. 

\emph{Ellipsoidal Variations:} Because of the planet's small semi-major axis and inflated radius, it is likely that it ---and possibly its host star--- are ellipsoidal in shape rather than spherical. This leads to changes in cross-sectional area throughout the planet's orbit. To good approximation, these variations are sinusoidal with a period half of the orbital period; the maxima occur at quadrature, when we are seeing the long axes of the star and planet, and minima at superior and inferior conjunction, when we are seeing the short axes of the two bodies.\footnote{This is a common approximation for ellipsoidal variations \citep[e.g.,][]{Faigler_2011}, but we discuss the exact expression in Section~5.3.1.} 

\cite{Li_2010}, \cite{Leconte_2011} and \cite{Budaj_2011} all predict that the projected area of WASP-12b should vary by approximately 10\% (peak-to-trough) due to its prolate shape, whether it is modeled as a prolate ellipsoid or a partially-filled Roche Lobe. Given the mid-infrared planet/star flux ratio of $F_{p}/F_{*}\approx4\times10^{-3}$ \citep{Campo_2011}, we expect to see ellipsoidal variations \emph{in the planet} at the level of $\Delta F/F_{*}\approx4\times10^{-4}$ (or a semi-amplitude of $2\times10^{-4}$). 

The presence of a massive companion should also produce ellipsoidal variations \emph{in the star}, as seen at optical wavelengths in the system HAT-P-7 \citep{Welsh_2010}. Using the expressions given in \cite{Faigler_2011}, we estimate the semi-amplitude of these variations to be $\sim4\times10^{-5}$ at all wavelengths. 
We therefore expect that at thermal wavelengths, the ellipsoidal variations of the system should be dominated by the shape of the planet and not that of its host star. 

We experimented with different functional forms for the planet's phase variations in an effort to reduce correlations between astrophysical variables.  Our best model in this regard is:  
\begin{equation}
\frac{F_{p}}{F_{*}} = \langle\frac{F_{p}}{F_{*}}\rangle + A_{\rm therm}\cos(\alpha-\alpha_{\rm max}) - A_{\rm ellips}\cos(2\alpha),
\end{equation}
where $A_{\rm therm}$ and $A_{\rm ellips}$ are the semi-amplitudes of diurnal and ellipsoidal phase variations, respectively, and $\alpha$ is the phase angle ($\alpha=0$ at superior conjunction, $\alpha=\pi$ at inferior conjunction). 

The secondary eclipse depth is related to the model variables by:
\begin{equation}
\frac{F_{\rm day}}{F_{*}} =  \langle\frac{F_{p}}{F_{*}}\rangle + A_{\rm therm}\cos\alpha_{\rm max} - A_{\rm ellips},
\end{equation}
and to first order the associated uncertainties can be propogated as: 
 \begin{eqnarray}
\sigma_{F_{\rm day}/F_{*}}^{2} = & \sigma_{\langle F_{p}/F_{*}\rangle}^{2} + \cos^{2}\alpha_{\rm max} \sigma_{A_{\rm therm}}^{2} \nonumber\\
&+ A_{\rm therm}^{2}\sin^{2}\alpha_{\rm max} \sigma_{\alpha_{\rm max}}^{2}+\sigma_{A_{\rm ellips}}^{2}.
\end{eqnarray}
In practice, this is an overestimate of uncertainty, because $\langle F_{p}/F_{*}\rangle$ and $A_{\rm therm}$ are anti-correlated.

\subsection{Detector Response Model}
\begin{figure*}[htb]
\begin{center}$
\begin{array}{cc}
\includegraphics[width=84mm]{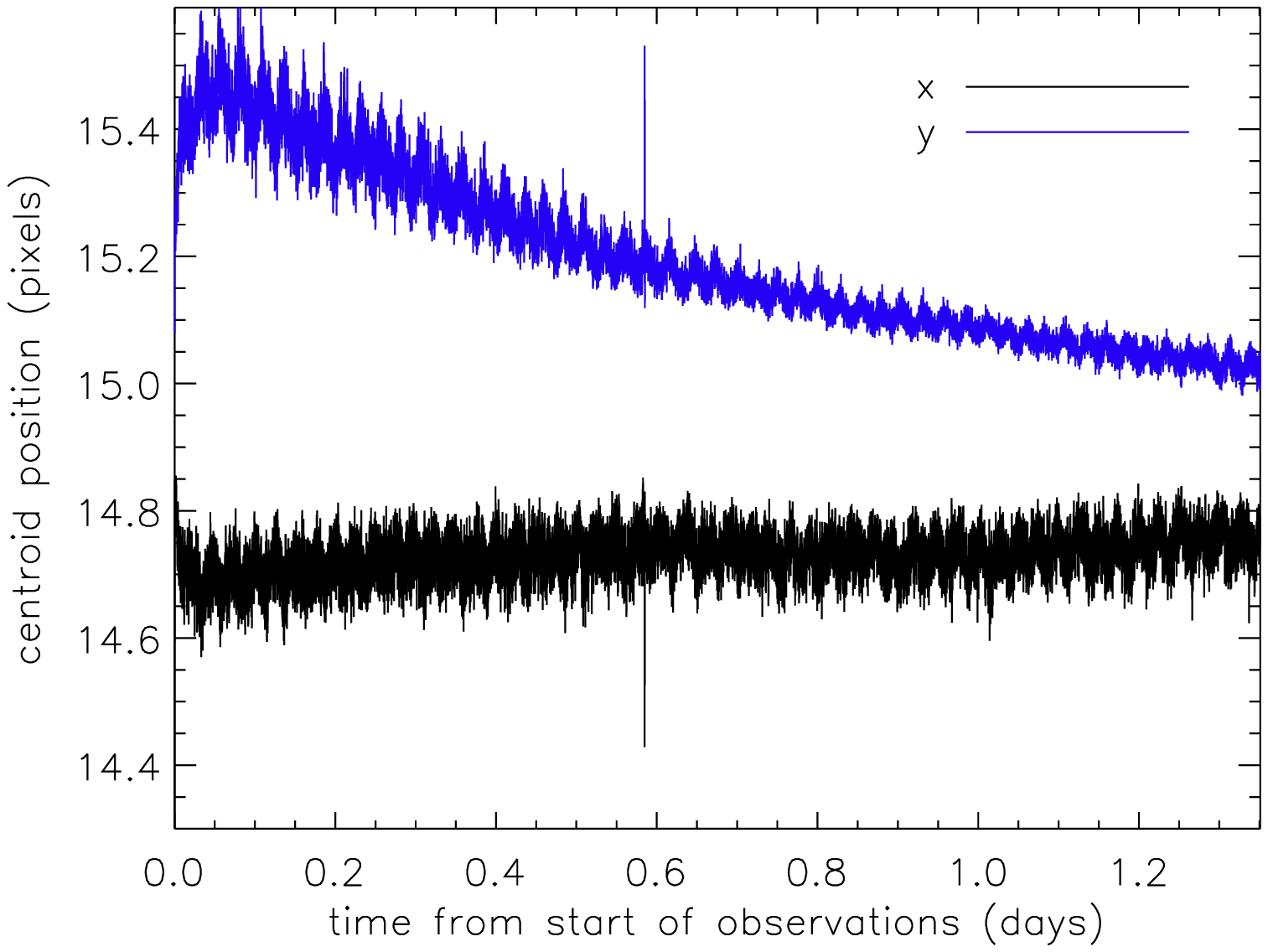} & \includegraphics[width=84mm]{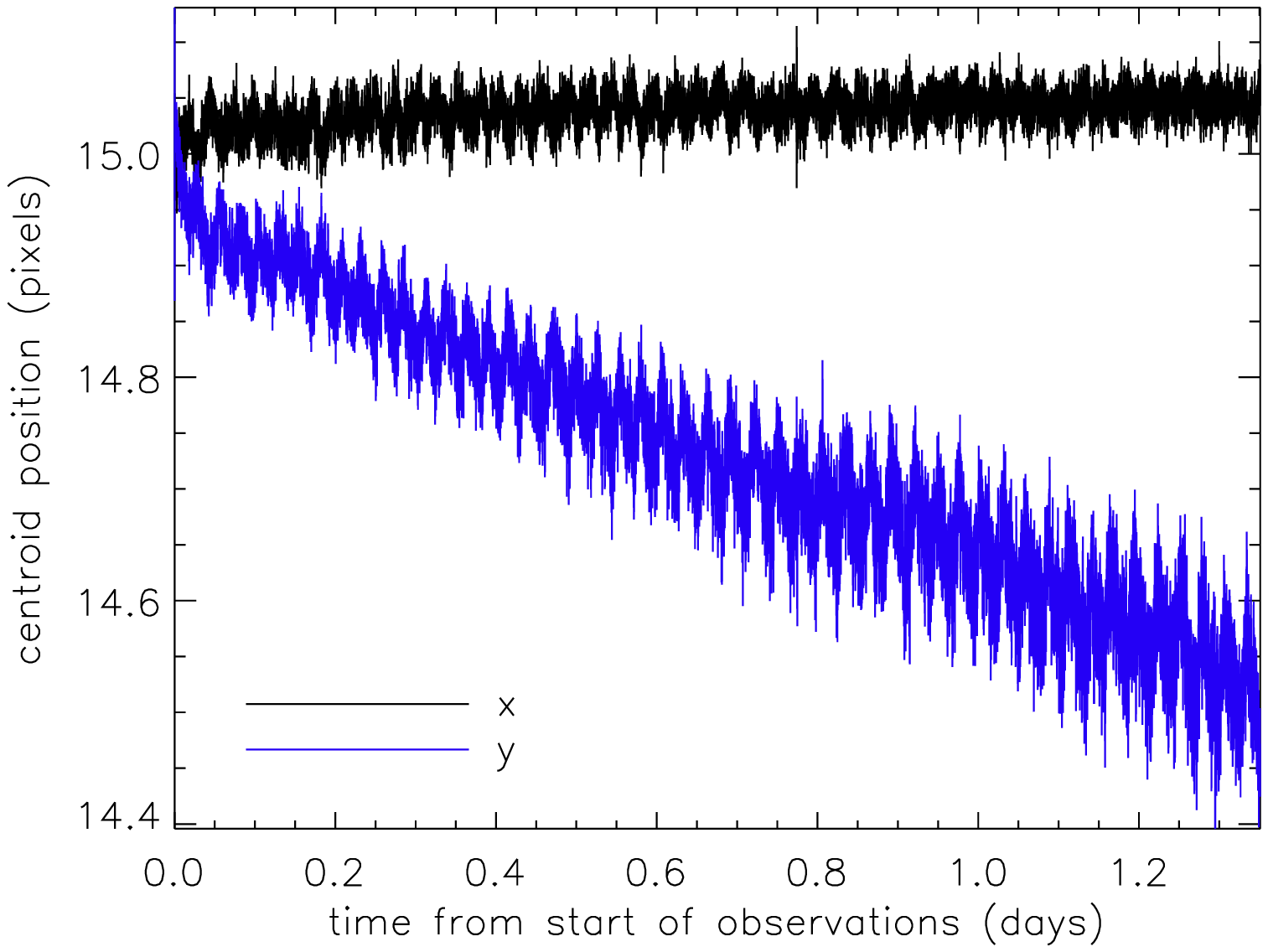} \\
\includegraphics[width=84mm]{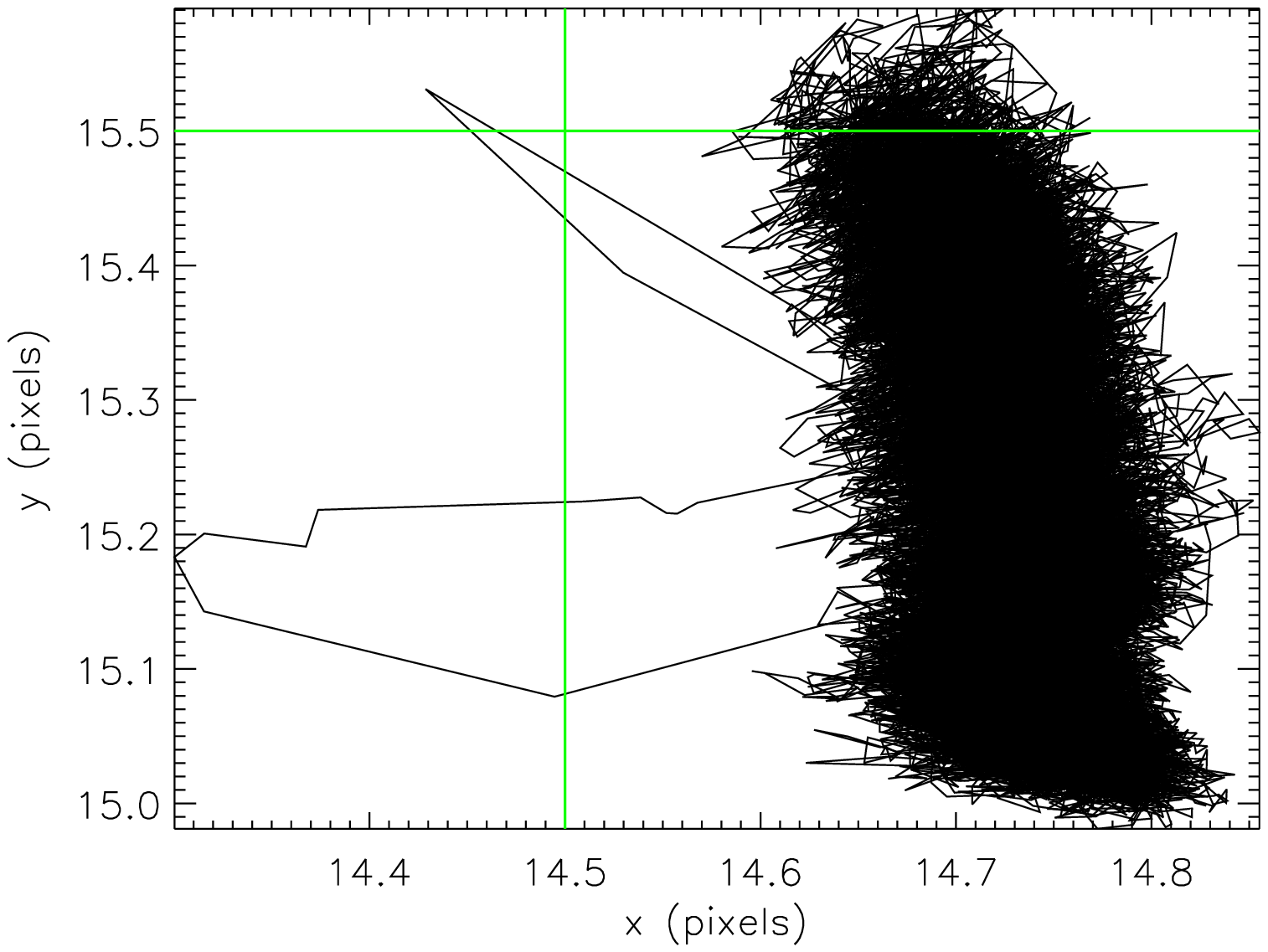} & \includegraphics[width=84mm]{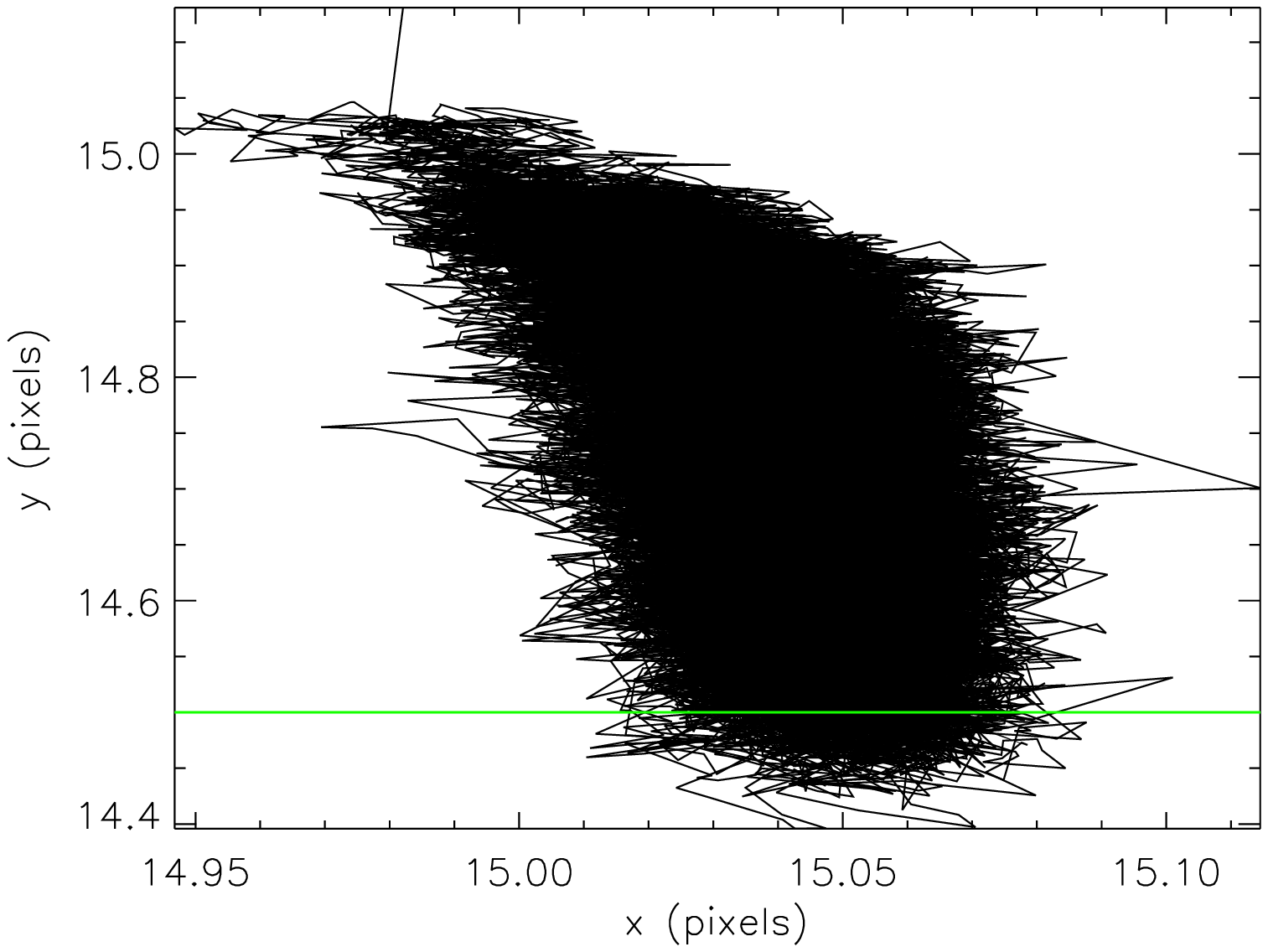}\\
\includegraphics[width=84mm]{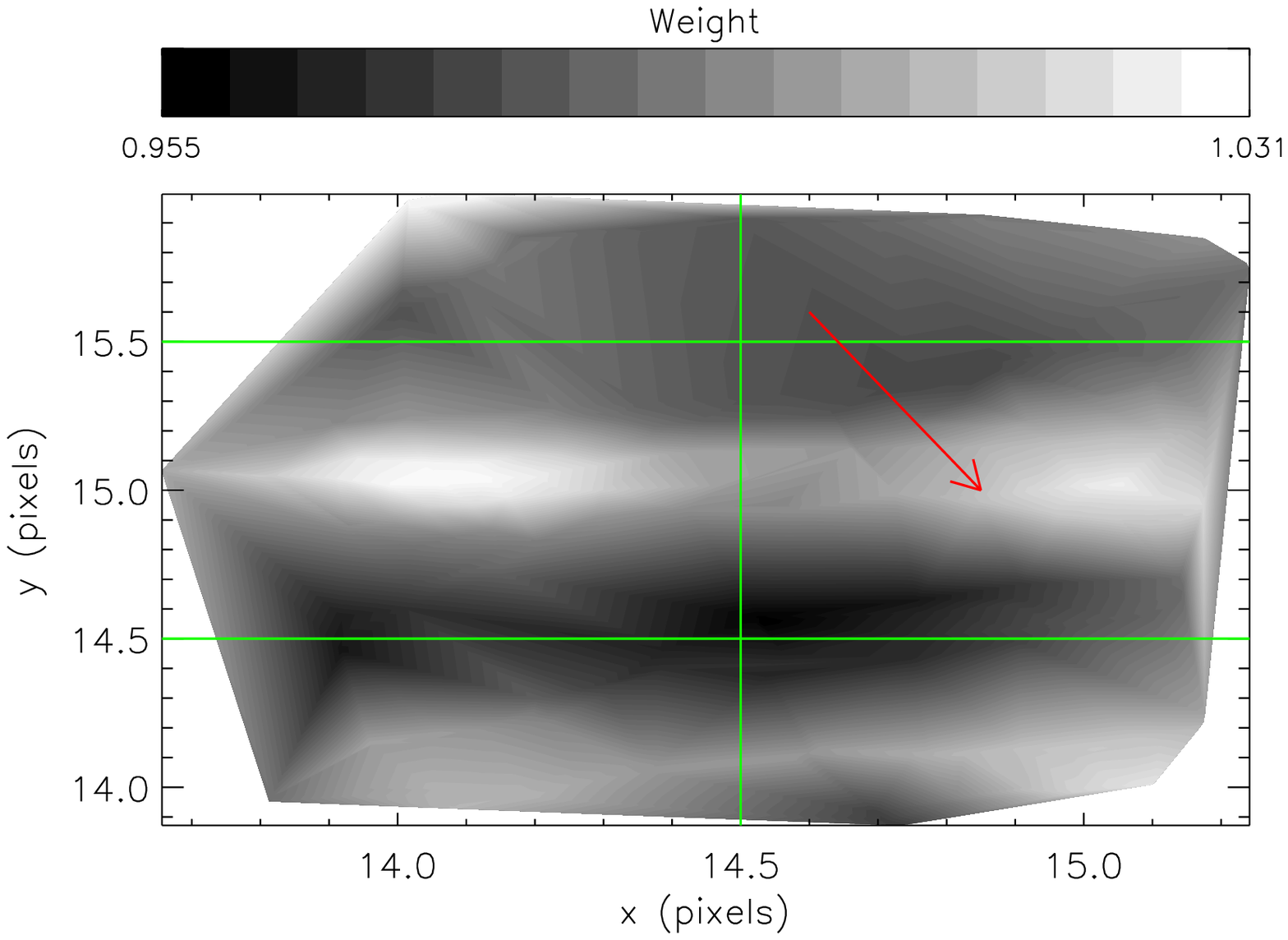} & \includegraphics[width=84mm]{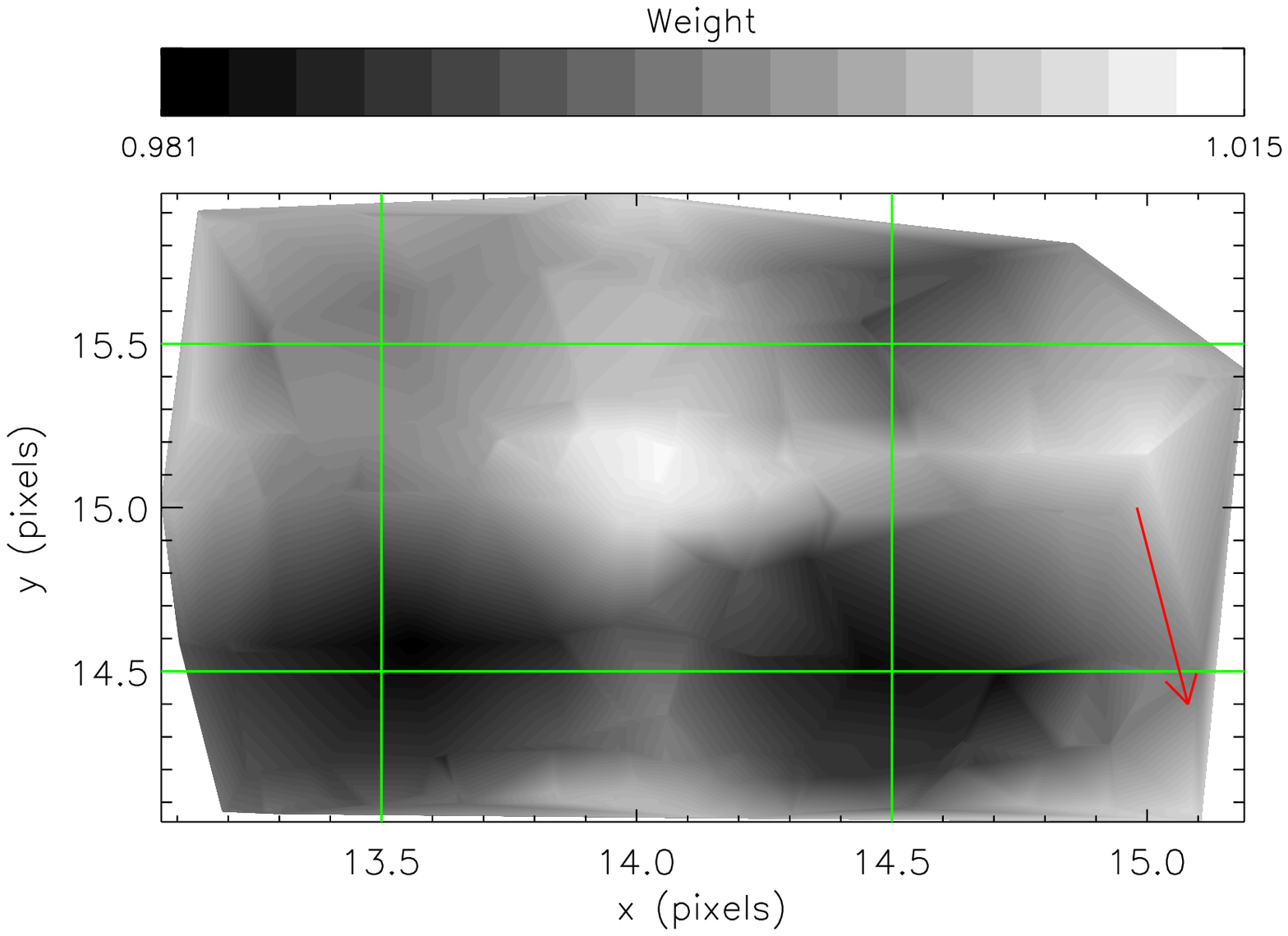}
\end{array}$ 
\end{center}
\caption{Point-spread-function (PSF) centroid movement at 3.6 $\mu$m (left) and 4.5 $\mu$m (right). The top panels shows the jitter and drift of the PSF centroid. The second panels show the two-dimensional wander of the centroid. The bottom panels shows the intra-pixel sensitivity map for the central regions of the detector, constructed by applying the \cite{Ballard_2010} point-by-point decorrelation to our mapping data (see Section~\ref{mapping}), with red arrows marking the approximate drift of the PSF over the course of our observations. The green lines show pixel edges.}
\label{ipsv}
\end{figure*}

IRAC channels 1 and 2 exhibit well-known intra-pixel sensitivity variations (IPSVs): photons hitting certain parts of a pixel lead to more electron counts than photons hitting other parts of the same pixel \citep[e.g.,][]{Charbonneau_2005, Morales_2006}.\footnote{This is entirely different than \emph{inter}-pixel sensitivity variations, which should have been largely calibrated out by flat-fielding. In any case, a scheme that corrects IPSVs implicitly corrects inter-pixel sensitivity variations as well.} In general, the sensitivity to photons is lowest near the edges of a pixel and greatest near its center (see bottom panels of Figure~\ref{ipsv}). On its own, IPSVs would not be a problem for precision time-resolved relative photometry. But over the course of observations, the point spread function (PSF) of the target star moves on the detector.  Even though the PSF spans many pixels, the IPSVs do not average out, because most of the flux falls in the PSF core.  

Since they are ultimately caused by changes in the PSF position, attempts to correct for IPSVs rely on accurate centroiding (described in Section~2).  The centroiding is shown in Figure~\ref{ipsv} for 3.6 $\mu$m (left) and 4.5 $\mu$m (right). The top panels show the centroid jitter and drift over the course of the observations; the second panels show the two-dimensional path of the centroid; the bottom panels show the intra-pixel sensitivity map of the four central pixels of the detector, constructed by applying the \cite{Ballard_2010} point-by-point decorrelation to our mapping data (see Section~\ref{mapping}).

Both $x$ and $y$ centroids exhibit fast jitter (period of roughly half an hour) with peak--trough amplitude of 0.05--0.1 pixels. This is the same jitter that used to have a period of roughly an hour: it is related to the heater cycling on \emph{Spitzer}.  The cycling frequency was doubled in fall 2010, which doubled the centroid jitter frequency and roughly halved the amplitude of the jitter.\footnote{ssc.spitzer.caltech.edu/warmmission/news/21oct2010memo.pdf} The smaller amplitude of the jitter is undoubtedly an improvement, while the higher frequency may or may not be a nuisance, depending on the duration of ingress/egress for a given planet.  In our data, the 3.6 $\mu$m flux exhibits clear 1\% peak-to-trough flux variations on the centroid-jitter timescale, while the 4.5 $\mu$m flux does not.

There is also a longer-term centroid drift, which is greatest in the $y$-direction: 0.5 pixel of motion over the $\sim$1 day observation at both 3.6 and 4.5 $\mu$m.  The $x$-direction shows almost no long-term drift (0.05 pixels, comparable to the faster jitter). Looking at the bottom panels of Figure~\ref{ipsv}, it is easy to understand why the IPSV's are worse at 3.6 $\mu$m than at 4.5 $\mu$m: at the shorter waveband, the PSF drifted up a steep slope in sensitivity from a pixel corner towards a center; at the longer waveband, the PSF contoured below a ridge in sensitivity.

The telescope takes a few hours to settle after pointing at a new target, resulting in larger PSF excursions for the first few data cubes of each lightcurve. It is difficult to correct for IPSVs in poorly-sampled regions of the detector, so we remove the first 0.05 days of both the 3.6 and 4.5 $\mu$m lightcurves (3.8\% of our data in each waveband).

The crux of the data reduction process is correcting for IPSVs, because our astrophysical signals (eclipses and phase variations) have an expected amplitude of 0.4\%, comparable to ---or smaller than--- these systematics.  We tried a variety of techniques, of which we describe the most promising below. We first present two methods for removing the IPSVs \emph{before} fitting our astrophysical model of the system. These techniques are attractive because they allow one to produce a systematics-corrected lightcurve independent of any astrophysical model assumptions.  We then present an attempt to simultaneously fit the IPSV and the astrophysical brightness variations.

\subsubsection{Gaussian Decorrelation}
\begin{figure*}[htb]
\begin{center}$
\begin{array}{cc}
\includegraphics[width=84mm]{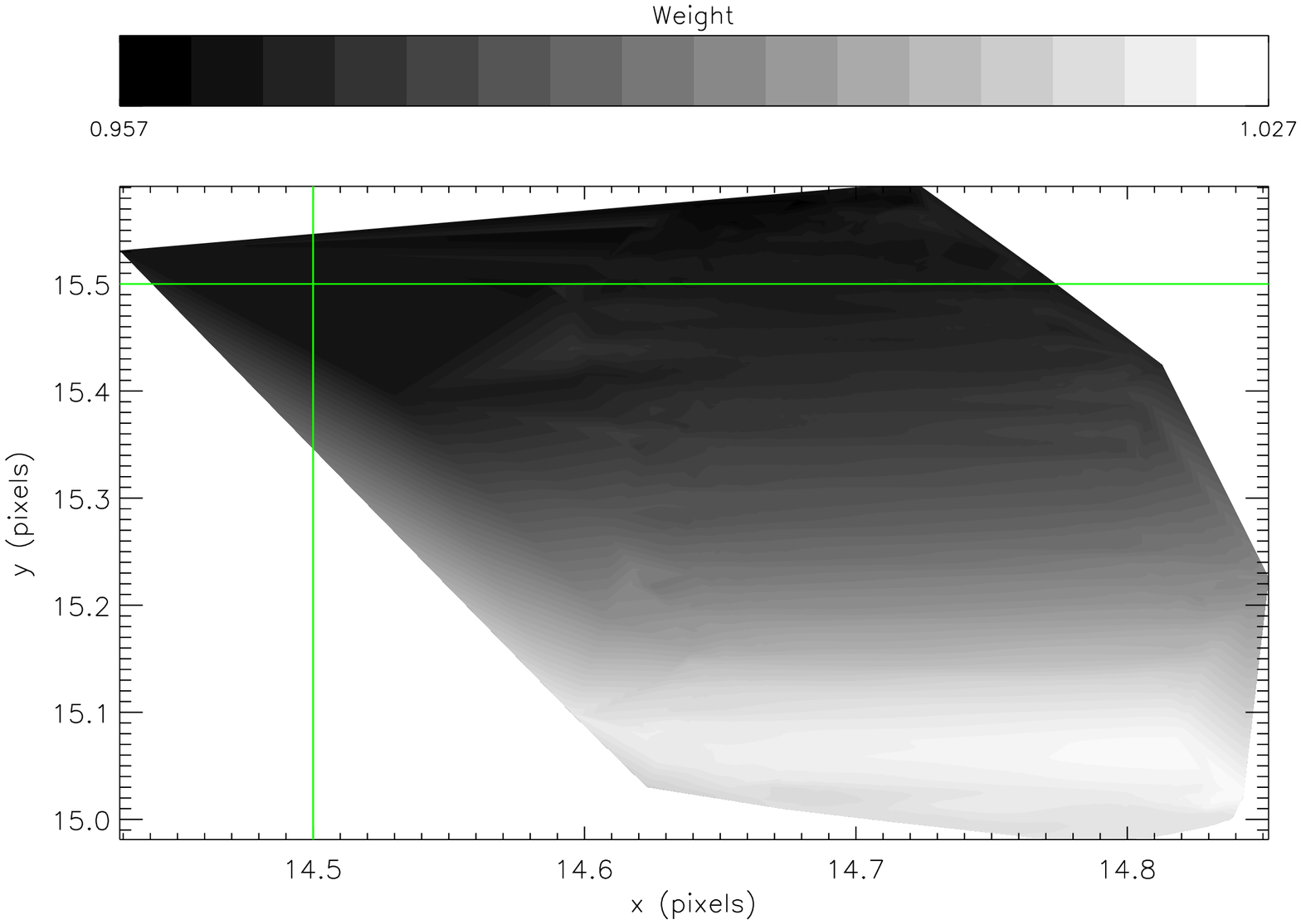} & \includegraphics[width=84mm]{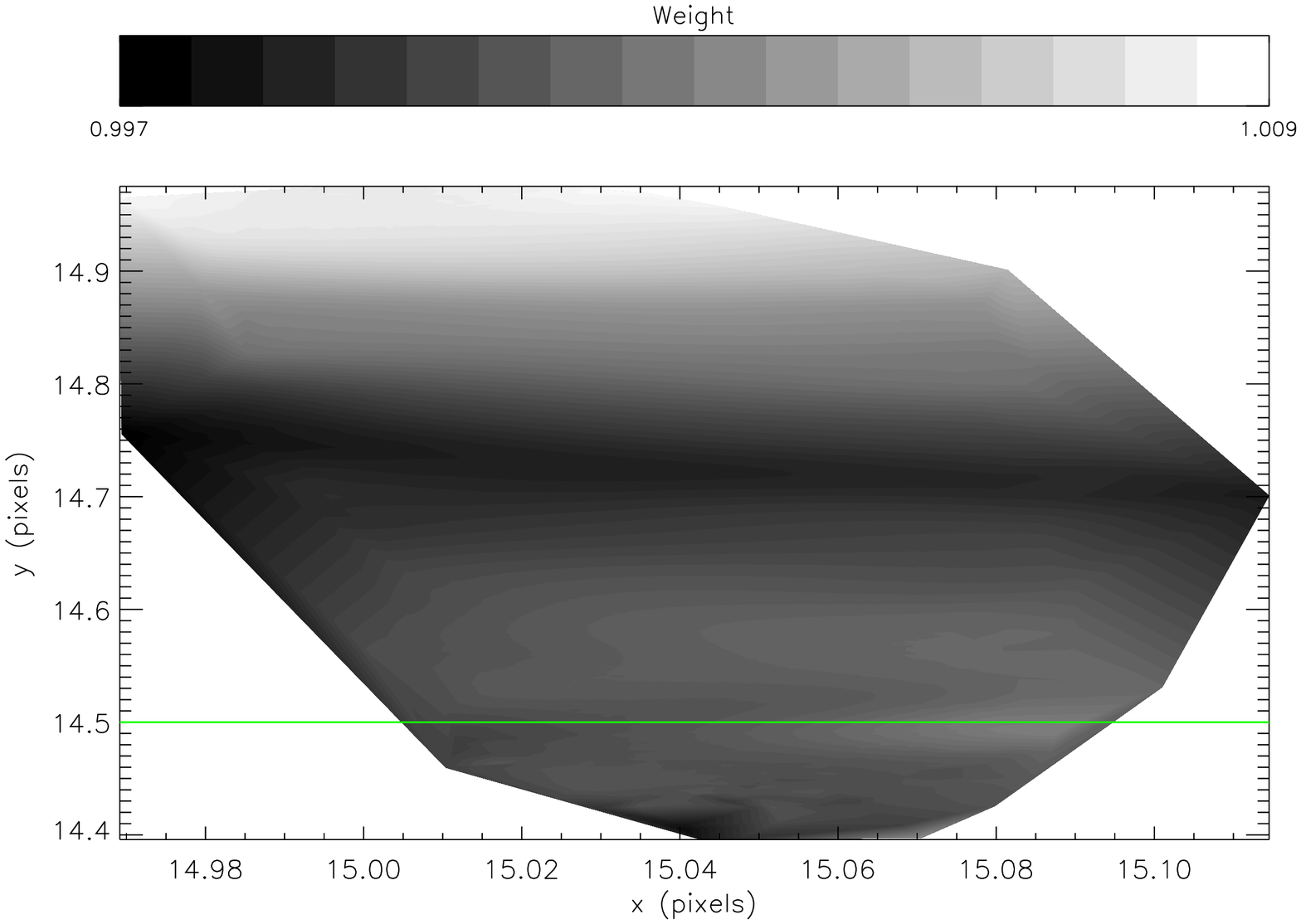} \\
\includegraphics[width=84mm]{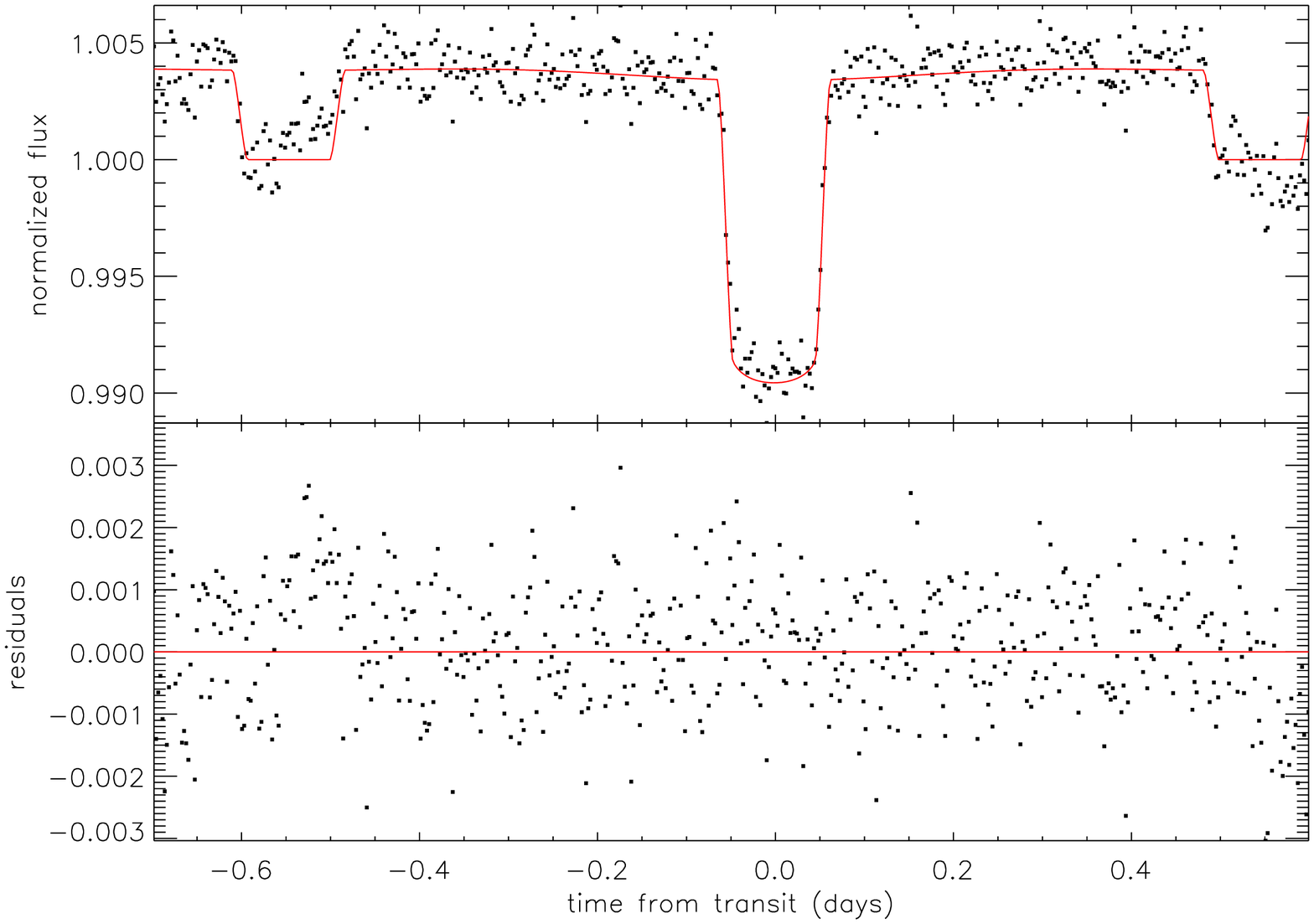} & \includegraphics[width=84mm]{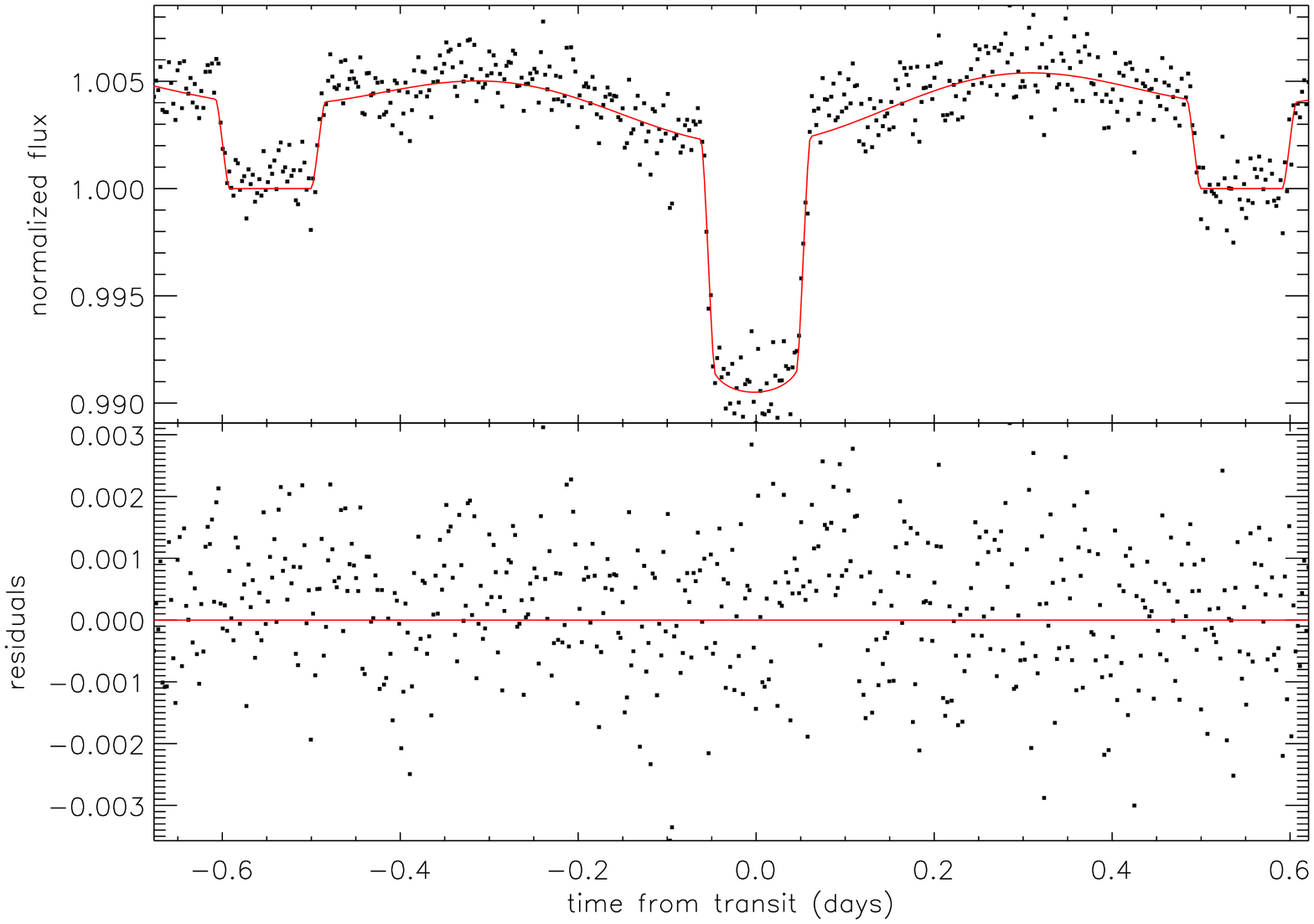}\\
\includegraphics[width=84mm]{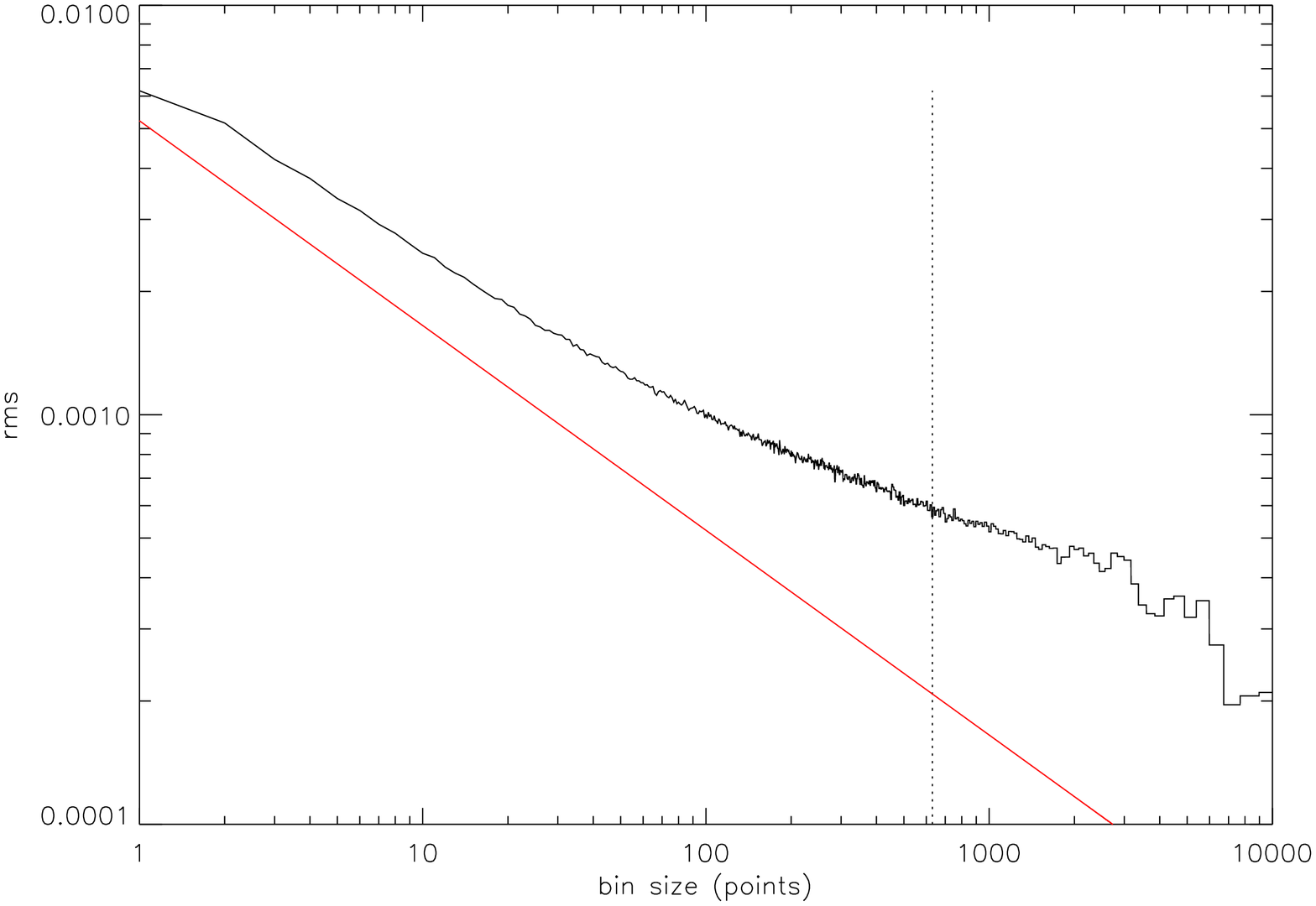} & \includegraphics[width=84mm]{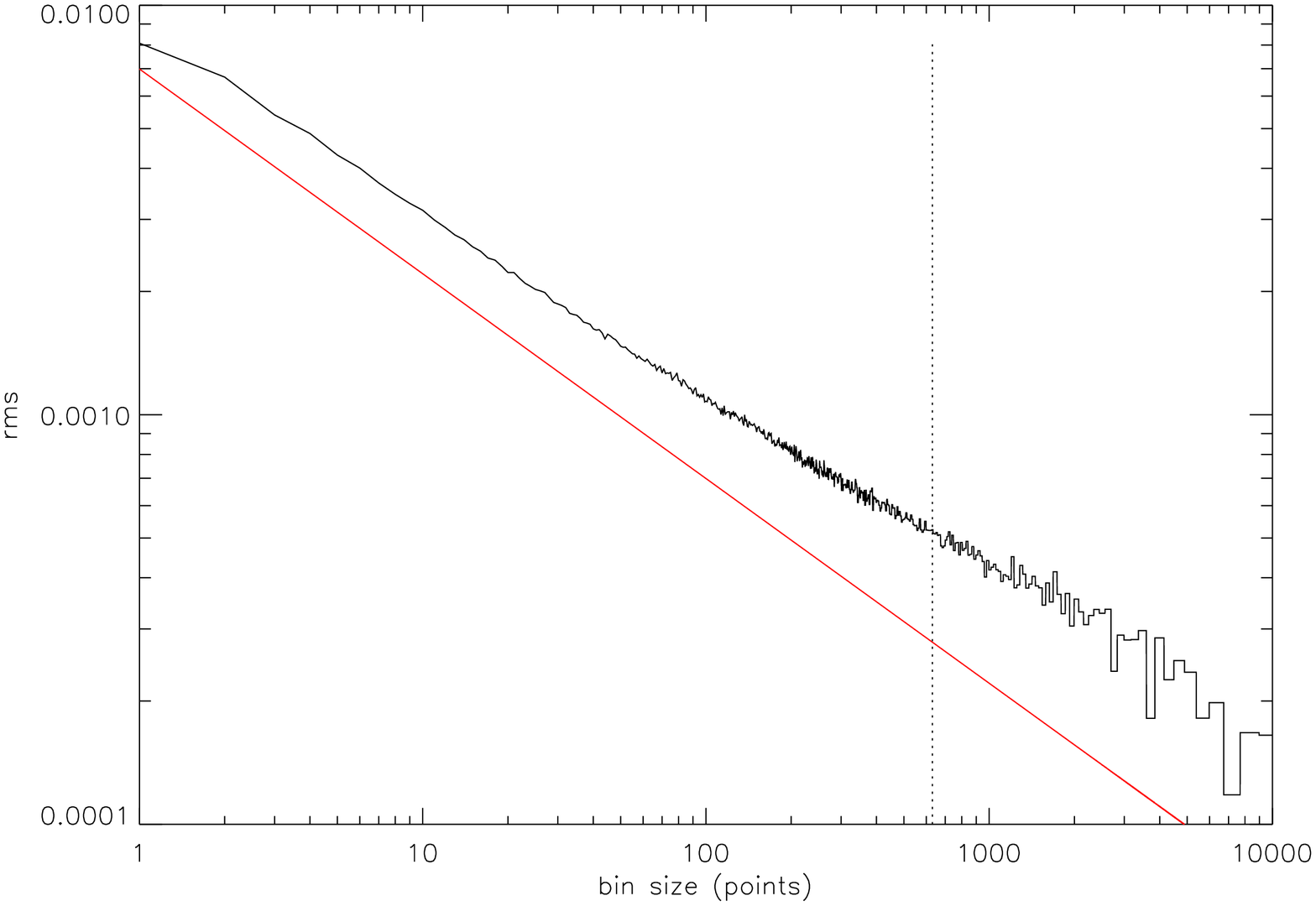}
\end{array}$
\end{center}
\caption{WASP-12b at 3.6 $\mu$m (left) and 4.5 $\mu$m (right), where we have corrected for intra-pixel sensitivity variations (IPSV's) using the \cite{Ballard_2010} point-by-point decorrelation, as described in Section~3.2.1. The upper panels show the IPSV map determined from our science data, with pixel edges shown in green.  The second panels show the corrected data with best-fit astrophysical model, and residuals. The bottom panels show the scatter in the residuals as a function of binning; the red line shows the photon noise limit; the vertical dotted line denotes the timescale of ingress/egress.}
\label{weight}
\end{figure*}

We follow \cite{Ballard_2010} in using point-by-point positions and fluxes to generate an intra-pixel sensitivity map with $x$ and $y$ Gaussian smoothing lengths of $\sigma_{x}$ and $\sigma_{y}$, respectively. \cite{Stevenson_2011} recently introduced a similar ---but faster--- method using bilinear interpolation. Since we are not attempting to iteratively fit the astrophysical and IPSV model, we use the simpler \cite{Ballard_2010} method.  We experimented with different smoothing lengths and chose the combinations that yield the smallest $\chi^{2}$ value for the final model fit; the astrophysical parameters are not very sensitive to changes in the smoothing length. At 3.6 $\mu$m we use $\sigma_{x}=0.017$ and $\sigma_{y}=0.0043$, as in \cite{Ballard_2010}; at 4.5 $\mu$m we use $\sigma_{x}=\sigma_{y}=0.015$. The resulting pixel maps are shown in the top panels of Figure~\ref{weight}.

We then divide the raw photometry by the weight function and fit our astrophysical model. The second panels of Figure~\ref{weight} show the corrected data with best-fit astrophysical model, and residuals. The bottom panels show the scatter in the residuals as a function of binning, along with a red line indicating the Gaussian-noise limit of root-mean-squared scatter scaling as $\sqrt{N}$. The normalization of this theoretical curve is based on the Poisson error for our electron counts (see first the raw photometry in Figure~\ref{wasp_12_raw}).  The best-fit astrophysical parameters are listed in Tables~\ref{best_fit_ch1} (3.6 $\mu$m) and \ref{best_fit_ch2} (4.5 $\mu$m).

\subsubsection{Mapping Data} \label{mapping}
The purpose of the mapping data was to deliberately map the central four pixels of the detector by scanning over them in 0.2-pixel and 0.1-pixel increments in the $x$- and $y$-directions, respectively. Since these observations are much shorter than the planet's orbital time, and were scheduled to avoid transits or eclipses, the changes in flux are in principle entirely due to the centroid position on the detector. 

In practice, the 3.6 $\mu$m lightcurve observations ended approximately 3.8 minutes before the end of eclipse egress, and the 3.6 $\mu$m mapping observations immediately followed. We therefore remove the first 4 minutes (0.003~days) of the 3.6 $\mu$m mapping data.

We use the mapping data centroids and fluxes to generate a weight map at the locations of the science centroids, again following \cite{Ballard_2010}. We adopt larger smoothing kernels set by the Nyquist sampling frequency of the regularly-spaced mapping centroids ($\sigma_{x}=0.1, \sigma_{y}=0.05$). We then use this weight function to correct the science lightcurve as above. 

Using the mapping data to generate pixel maps has the advantage that we are not self-calibrating our science data, and hence are not liable to throw the baby out with the bath water. On the other hand, it does not successfully remove the systematics: the model fits are far worse than either the Gaussian decorrelation discussed above, or the polynomial models discussed below (the discrepancy in $\chi^{2}$ is a factor of $\sim10$ at 3.6 $\mu$m and $\sim4$ at 4.5 $\mu$m). This means that the IPSV must have fine spatial structure \citep[as seen by][]{Ballard_2010}, and/or some additional flux- or time-dependence. 

\subsubsection{Polynomial IPSV Model} \label{simul}
Here we model the intra-pixel sensitivity variations as a polynomial in the centroid $x$ and $y$. We simultaneously fit our astrophysical model and the IPSVs by treating the $x$ and $y$ centroids as independent variables in our function.  
We model the IPSV's as:
\begin{equation}
\frac{F_{\rm obs}}{F_{\rm astro}} =  \frac{1 + \sum_{i=1}^{n} \left[a_{i}(x-\bar{x})^{i} + b_{i}(y-\bar{y})^{i}\right]}{\langle1 + \sum_{i=1}^{n} \left[a_{i}(x-\bar{x})^{i} + b_{i}(y-\bar{y})^{i}\right]\rangle},
\end{equation}
where $F_{\rm obs}$ is the observed flux, $F_{\rm astro}$ is the astrophysical model, and $\bar{x}$ and $\bar{y}$ are the mean centroid positions. Formally, cross-terms are necessary to describe an arbitrary two-dimensional function, but we find that including cross-terms does not significantly improve the $\chi^{2}$ or affect the astrophysical parameters.  This is a testament to the fact that the bulk of the PSF motion is in the $y$-direction.

We experiment with polynomials up to sixth-order. To test whether each additional pair of coefficients (one each for $x$ and $y$) improved the fit, we use the Bayesian Information Criterion \citep[BIC;][]{Schwarz_1978}, which imposes a penalty term on the $\chi^{2}$ for additional free parameters: $BIC = \chi^{2} + k\ln N$, where $k$ is the number of free parameters and $N\approx52000$ is the number of data. Since $\ln(52000)\approx 11$, an additional parameter is acceptable if it improves the $\chi^{2}$ by at least 11. The BIC for our 3.6 $\mu$m data improved with the addition of parameters up to and including fifth order. The BIC for our 4.5 $\mu$m data was not improved by the addition of terms beyond cubic.   

Unlike some previous studies, we do not include a linear ramp \emph{in time}.  Since the bulk of the PSF motion is a monotonic drift in the y-direction, we found the ramp in time to be highly correlated with the linear and quadratic coefficients of the $y$ sensitivity, and the fit was not significantly improved.\footnote{We do include a linear ramp in time when performing isolated fits to transits or eclipses, since 1) those shorter time-series do not provide enough leverage to properly fit the IPSV, and 2) the linear ramp can act as a proxy for the thermal phase variations, which are not explicitly fit for in these cases.}

\begin{figure*}[htb]
\begin{center}
\includegraphics[width=175mm]{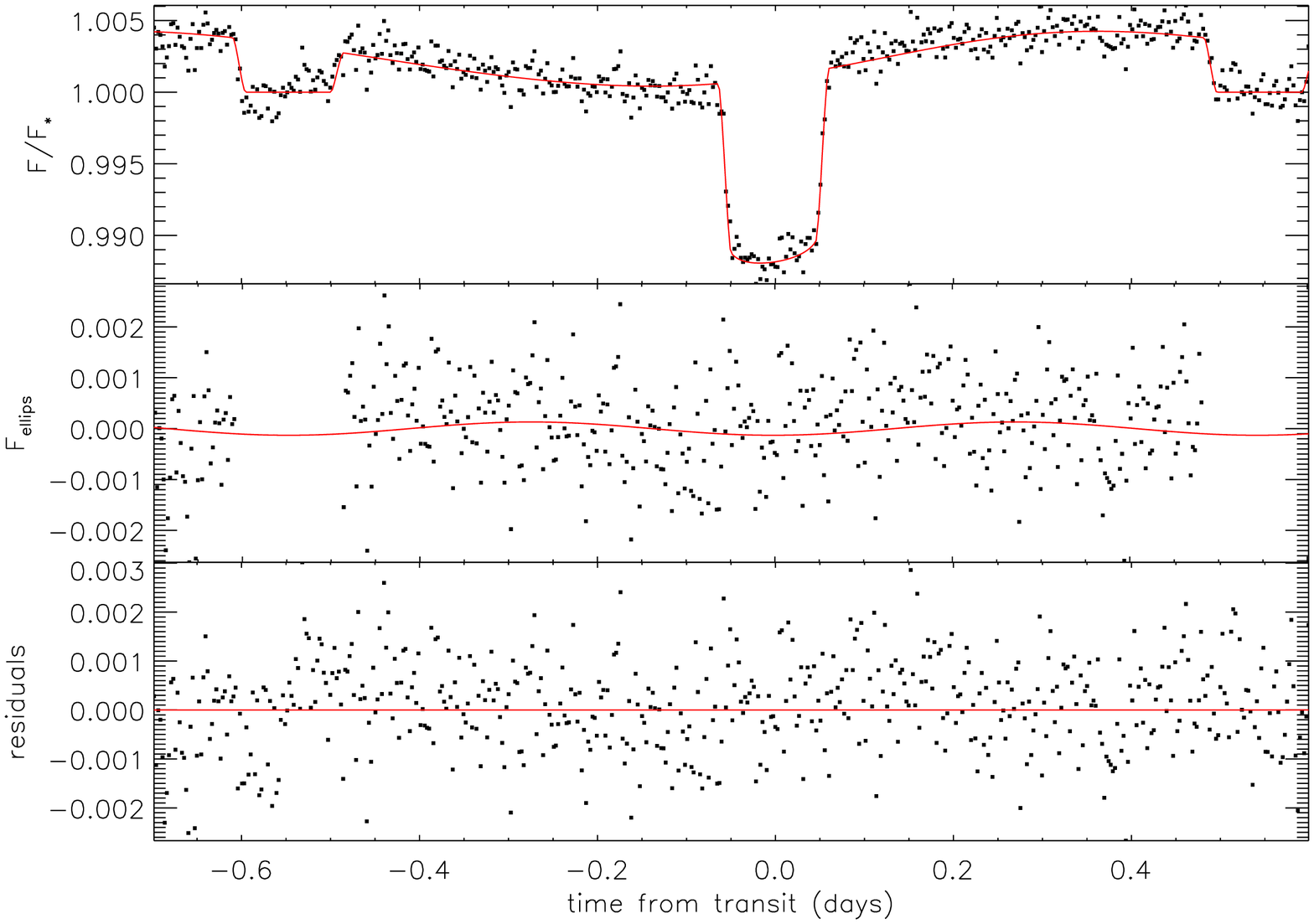}
$\begin{array}{cc}
\includegraphics[width=84mm]{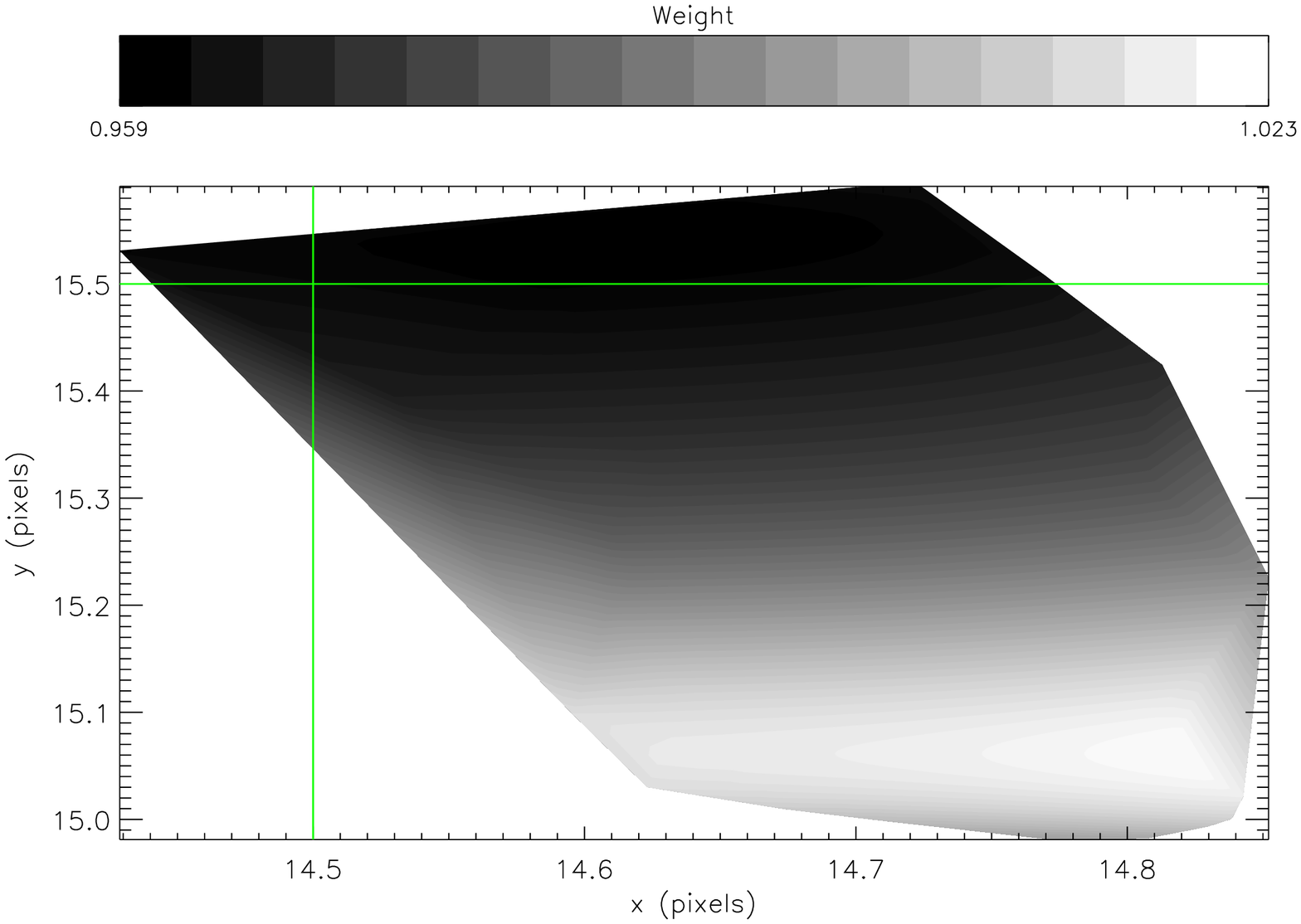}&\includegraphics[width=84mm]{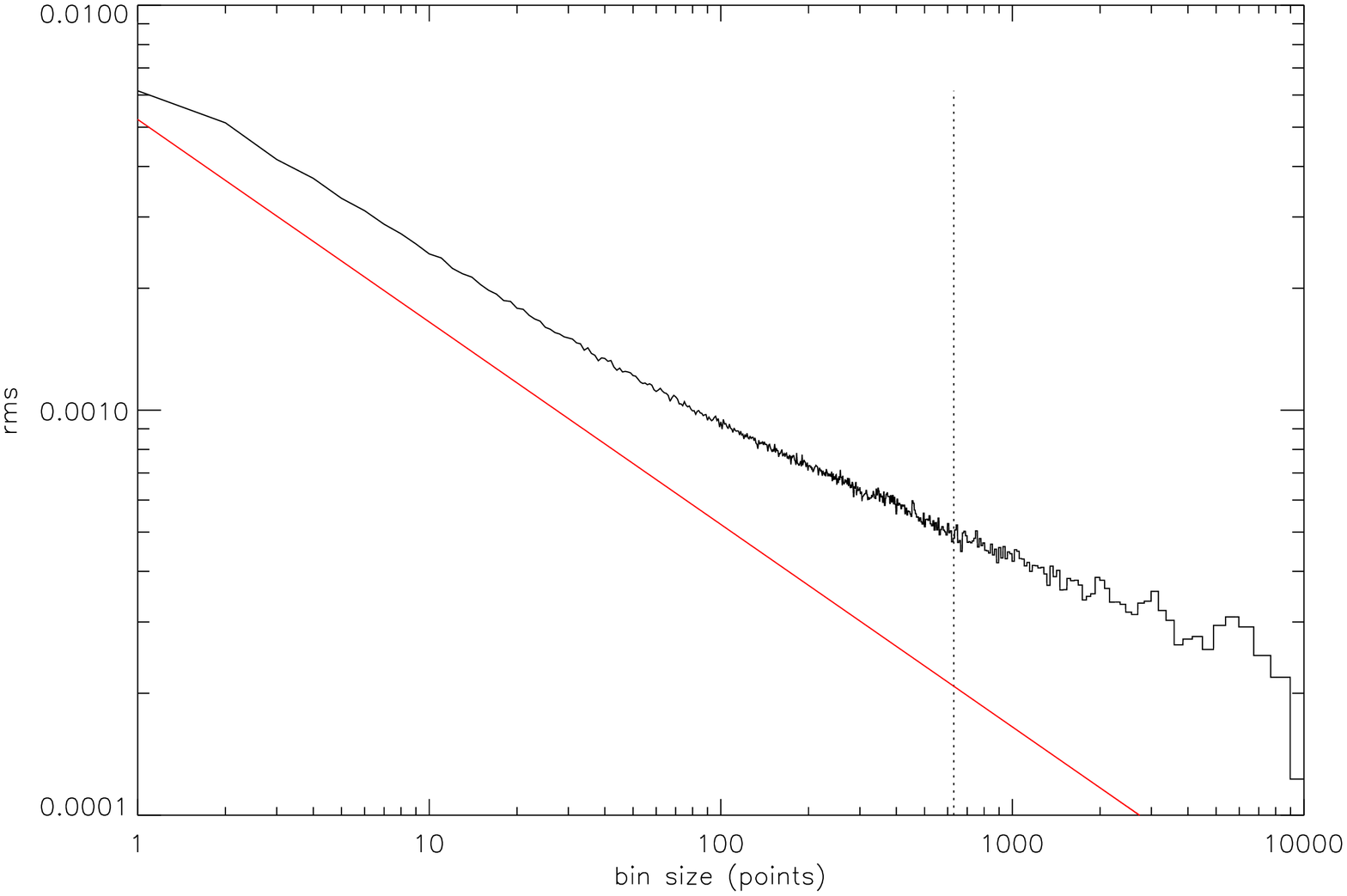}
\end{array}$
\end{center}
\caption{WASP-12b at 3.6 $\mu$m, where we have treated the IPSVs as a polynomial function in both $x$ and $y$ centroid, as described in Section~3.2.3.  The top panel shows: the systematics-corrected lightcurve and best-fit astrophysical model (top inset), the residuals after subtracting the best-fit transit, eclipse and thermal phase model, along with the best-fit ellipsoidal variations model (middle inset), and the residuals after removing ellipsoidal variations (bottom inset). Ellipsoidal variations of the planet do not affect in-eclipse data since the planet is hidden from view; hence we remove the in-eclipse data from this panel. The bottom-left panel shows the weight function used to correct the data, with pixel edges shown in green. The bottom-right panel shows the scatter in the residuals as a function of binning; the red line shows the photon noise limit; the vertical dotted line denotes the timescale of ingress/egress.}
\label{wasp_12_slope_ch1}
\end{figure*}

\begin{figure*}[htb]
\begin{center}
\includegraphics[width=175mm]{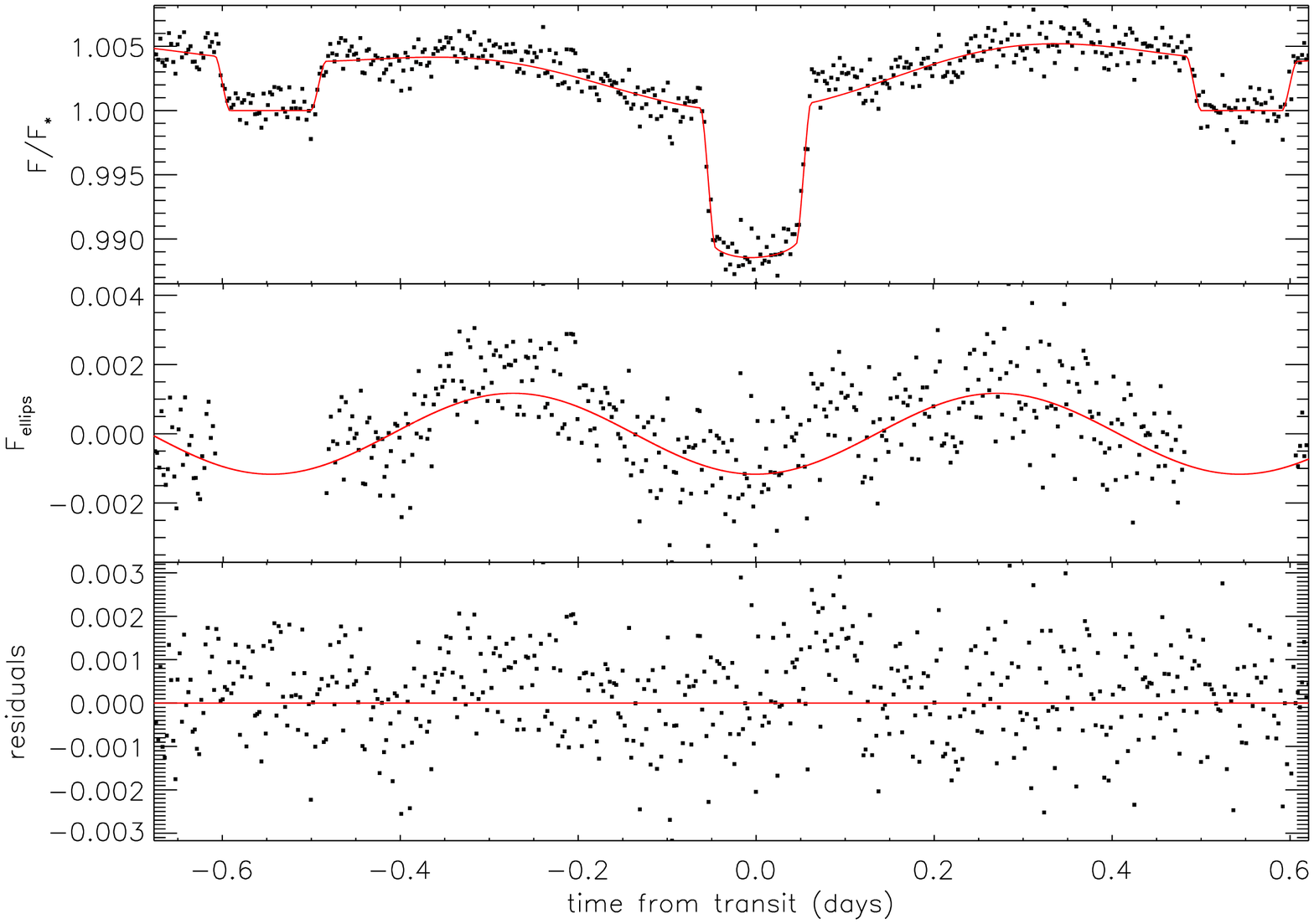}
$\begin{array}{cc}
\includegraphics[width=84mm]{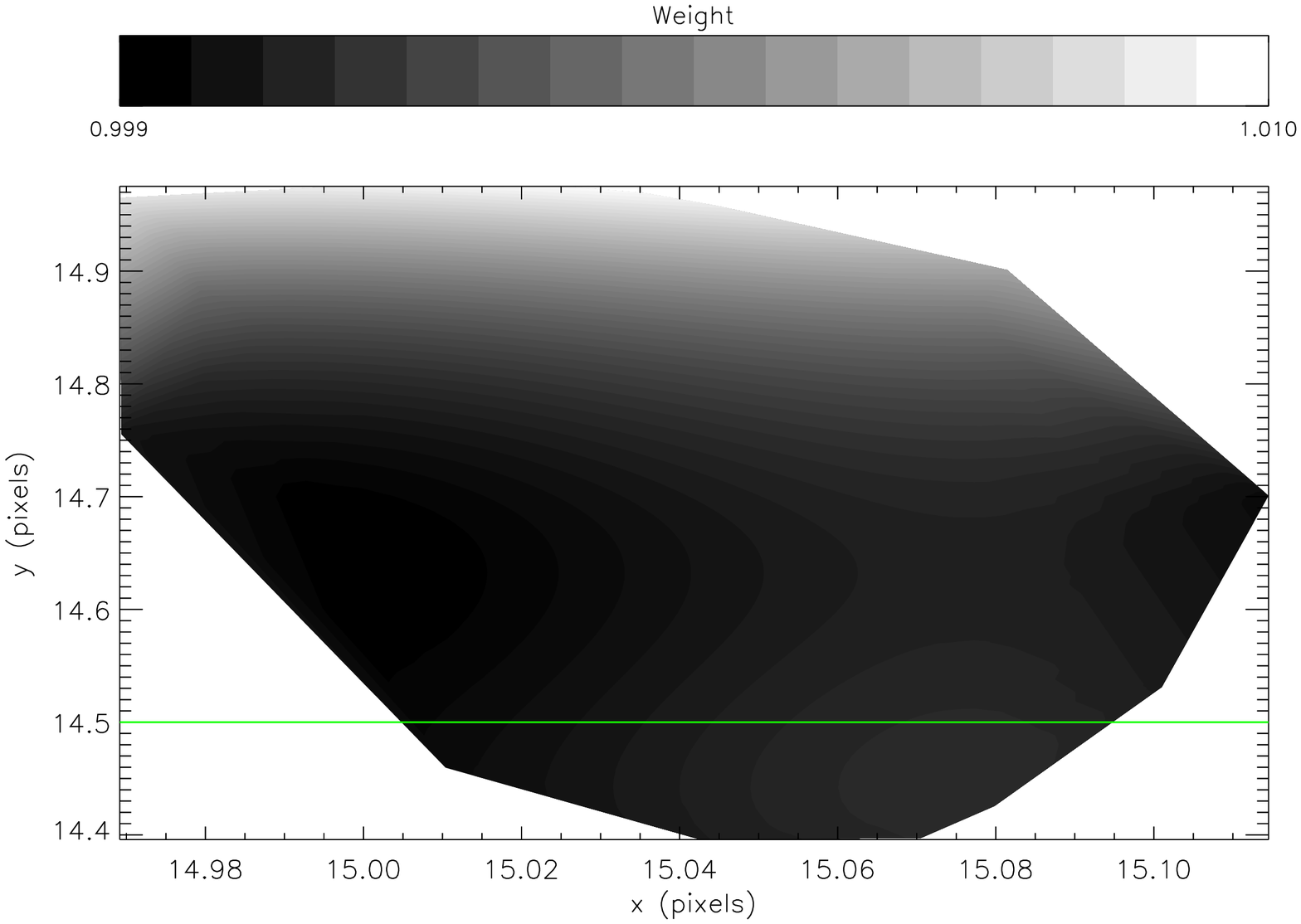}&\includegraphics[width=84mm]{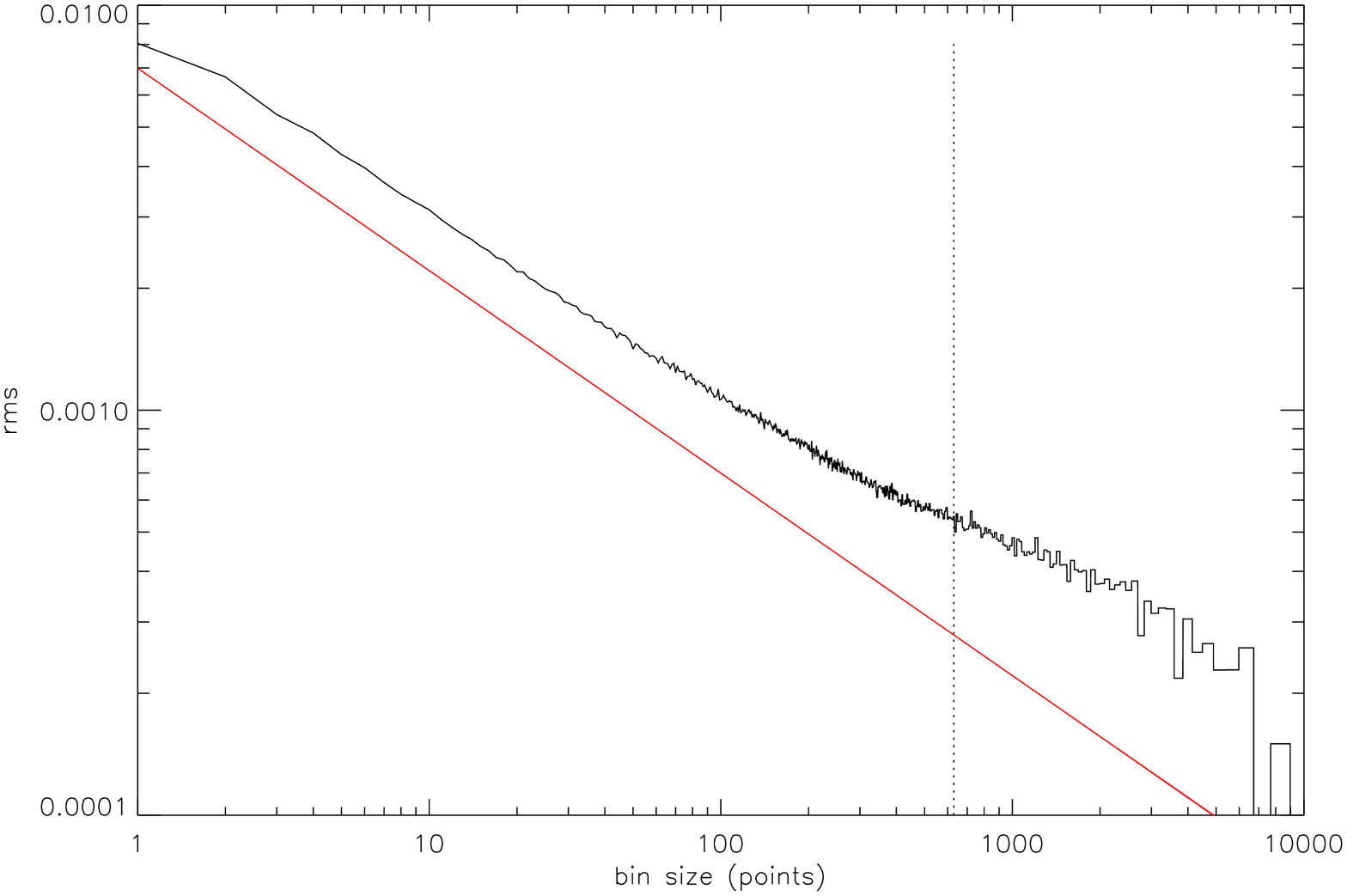}
\end{array}$
\end{center}
\caption{WASP-12b at 4.5 $\mu$m, where we have treated the IPSVs as a polynomial function in both $x$ and $y$ centroid, as described in Section~3.2.3.  The top panel shows: the systematics-corrected lightcurve and best-fit astrophysical model (top inset), the residuals after subtracting the best-fit transit, eclipse and thermal phase model, along with the best-fit ellipsoidal variations model (middle inset), and the residuals after removing ellipsoidal variations (bottom inset).  Ellipsoidal variations of the planet do not affect in-eclipse data since the planet is hidden from view; hence we remove the in-eclipse data from this panel. The bottom-left panel shows the weight function used to correct the data, with pixel edges shown in green. The bottom-right panel shows the scatter in the residuals as a function of binning; the red line shows the photon noise limit; the vertical dotted line denotes the timescale of ingress/egress.}
\label{wasp_12_slope_ch2}
\end{figure*}

We show the resulting fits in Figures~\ref{wasp_12_slope_ch1} and \ref{wasp_12_slope_ch2}. For each figure, the top panel shows: the systematics-corrected lightcurve and best-fit astrophysical model (top inset), the residuals after subtraction of the best-fit thermal phases, along with the best-fit ellipsoidal variations model (middle inset), and the residuals after removing ellipsoidal variations (bottom inset). (Ellipsoidal variations of the planet do not affect in-eclipse data since the planet is hidden from view; hence we remove the in-eclipse data from this panel.) The bottom-left panel shows the weight function used to correct the data. The bottom-right panel shows the scatter in the residuals as a function of binning; the red line shows the photon noise limit; the vertical dotted line denotes the timescale of ingress/egress.

\begin{deluxetable*}{lrrrrrrrr}
\tabletypesize{\scriptsize}
\tablecaption{3.6 micron Parameters \label{best_fit_ch1}}
\tablewidth{0pt}
\tablehead{
\colhead{Calibration Method} & \colhead{$\chi_{\rm R}^{2}$} & \colhead{$b$} & \colhead{$a/R_{*}$}&\colhead{$(R_{p}/R_{*})^{2}$} & \colhead{$F_{\rm day}/F_{*}$} & \colhead{$2A_{\rm therm}$} & \colhead{$\alpha_{\rm max}$} & \colhead{$A_{\rm ellips}$}}
\startdata
Point-by-Point Decorrelation & 1.392$^{a}$ & 0.5(1)& 2.8(2)&0.0125(4) &0.0038(4)& 0.0004(3) & 0(29)$^{\circ}$ & $1(1)\times10^{-4}$ \\
{\bf $^{b}$Polynomial in $x$ and $y$} &1.384$^{a}$ & 0.3(2)& 3.1(2) &0.0123(3) &0.0033(4)  & 0.0038(6) & -53(7)$^{\circ}$& $1(1)\times 10^{-4}$
\enddata
\tablenotetext{a}{When comparing these values, it is worth remembering that with approximately 52,000 degrees of freedom, a model is significantly better if it improves $\chi_{R}^{2}$ by at least 0.004.}
\tablenotetext{b}{Our fiducial analysis.}
\end{deluxetable*}

\begin{deluxetable*}{lrrrrrrrr}
\tabletypesize{\scriptsize}
\tablecaption{4.5 micron Parameters \label{best_fit_ch2}}
\tablewidth{0pt}
\tablehead{
\colhead{Calibration Method} & \colhead{$\chi_{\rm R}^{2}$} & \colhead{$b$} & \colhead{$a/R_{*}$}& \colhead{$(R_{p}/R_{*})^{2}$} &\colhead{$F_{\rm day}/F_{*}$} & \colhead{$2A_{\rm therm}$} & \colhead{$\alpha_{\rm max}$} & \colhead{$A_{\rm ellips}$}}
\startdata
Point-by-Point Decorrelation & 1.326$^{a}$ & 0.5(1)&2.9(2) &0.0112(4)&0.0039(3)& 0.0019(3) & -12(6)$^{\circ}$ & $1.1(1)\times10^{-3}$\\
{\bf $^{b}$Polynomial in $x$ and $y$} &1.324$^{a}$& 0.5(1)& 2.9(2) &0.0111(4)& 0.0039(3)& 0.0040(3) &-16(4)$^{\circ}$ & $1.2(2)\times10^{-3}$
\enddata
\tablenotetext{a}{When comparing these values, it is worth remembering that with approximately 52,000 degrees of freedom, a model is significantly better if it reduces $\chi_{R}^{2}$ by at least 0.004.}
\tablenotetext{b}{Our fiducial analysis.}
\end{deluxetable*}

\section{Model Fitting \& Error Analysis}
We use the IDL implementation of a Levenberg-Marquardt (L-M) $\chi^{2}$ gradient descent routine, \verb+MPFITFUN+, to find the best-fit model parameters.  The covariance matrix of the model parameters provides a first guess at the parameter uncertainties.

L-M or Markov Chain Monte Carlo (MCMC) error estimation depend on the photometric uncertainties: larger error bars on the data lead to larger uncertainties on the model parameters. For the initial fits, we optimistically set the error bars on our data at the Poisson limit, $1/\sqrt{N_{\rm counts}}$; this means that the reduced $\chi^{2}$ of our best-fit model fits is somewhat larger than 1 (see Tables~\ref{best_fit_ch1} and \ref{best_fit_ch2}), and it means that either L-M or MCMC will underestimate parameter uncertainties.  To alleviate this problem, we then scale the data uncertainties to give a reduced $\chi^{2}$, $\chi^{2}_{R}$, of unity (by multiplying uncertainties by the square-root of the best $\chi_{R}^{2}$). In the present case, this entails inflating the error bars by a constant factor of 10--20\%, depending on the waveband and model in question.  
Scaling photometric errors to obtain a reduced $\chi^{2}$ of unity 
renders the $\chi^{2}$ useless for comparing different models; we therefore quote the best reduced $\chi^{2}$ \emph{prior} to inflating the error bars.  After adjusting the photometric uncertainties, we normalize the lightcurve to the in-eclipse (star-only) value, and fix $F_{*}$ to unity in order to avoid correlations between $F_{*}$ and $\langle F_{p}/F_{*}\rangle$ in the final fits. 

It is well known that simply using the covariance matrix from the L-M fit does not provide a robust error estimate, so we estimate parameter uncertainties using a variety of other techniques.  We experimented with MCMC and Bootstrap Monte Carlo error estimates and found them to be slightly larger than ---but comparable to--- those from the L-M.  We found that our most conservative error estimates (typically larger by a factor of 2 or more) are obtained by considering the residuals of our best-fit model: either binning of residuals, or resampling of residuals using a prayer-bead Monte Carlo.  Throughout the manuscript we always adopt the largest uncertainty for a given parameter, but it is worth noting that even our most conservative error estimates may still be 15--30\% smaller than what one would obtain with a wavelet analysis \citep{Carter_2009}.

\subsection{Binning of Residuals}
\cite{Pont_2006} proposed a simple method to account for red noise by considering how the scatter in residuals decreases with bin size (bottom panels of Figures 3, 4 and 5).  At the left end of the plots, where we are considering point-to-point scatter in the residuals, the scatter is only 10--20\% greater than the Poisson counting limit shown in red ($\propto \sqrt{M/N(M-1)} \approx \sqrt{N}$, where $N$ is the number of observations per bin, and $M$ is the number of bins).  But the observed scatter does not follow the theoretical relation as the data are binned. The most important timescale for transit and eclipse parameter estimation is the duration of ingress/egress, which we denote by a vertical dotted line in those panels (21 minutes for WASP-12b).  The scatter on this timescale determines the accuracy we can expect to achieve for transit or eclipse depths. 

Following \cite{Winn_2007, Winn_2008}, we define the factor $\beta$ as the actual scatter (black line) divided by the theoretical Poisson limit (red line) on the 21 minute timescale (vertical dotted black line); our residuals have $\beta$ = 2--3. To account for red noise in parameter uncertainties, we simply inflate the L-M parameter uncertainties by $\beta$. (Inflating the photometric error bars by $\beta$ and recomputing the covariance matrix using L-M takes longer and produces slightly smaller parameter uncertainties.)  The binning of residuals method turns out to be the most conservative error estimate for transit- and eclipse-specific model parameters ($b$, $a/R_{*}$, $(R_{p}/R_{*})^{2}$, etc.).  

\subsection{Boot-Strap \& Prayer-Bead Monte Carlo}
We also estimate parameter uncertainties using two resampling techniques: boot-strap and prayer-bead Monte Carlos. Both of these techniques use the scatter in the residuals of our best fit model as an estimate of photometric uncertainty. In both cases the residuals are shifted in time and added back to the best fit model to produce a new instance of the lightcurve, which is then fit using the L-M.\footnote{Since we use L-M to find best-fit solutions, one might think that the prayer-bead and boot-strap analyses implicitly depend on photometric error estimates.  But in fact, the L-M algorithm settles on the same solution irrespective of error bars (within reason). The prayer bead and bootstrap parameter uncertainties are therefore effectively independent of the photometric uncertainties.} The standard deviation in the sequence of model parameters is our estimate of their $1\sigma$ uncertainty. Note that ---by construction--- resampling techniques cannot improve parameter estimates: the best-fit parameters are those determined by fitting the original time series. 

For the boot strap Monte Carlo, the residuals are randomly shuffled so ---much like the L-M and MCMC techniques--- the bootstrap error analysis is insensitive to the ordering of residuals.  Such error estimates will therefore only be accurate insofar as residuals are uncorrelated in time. Indeed, we found that error estimates from the bootstrap were comparable to those from the L-M or MCMC analyses.  

The prayer-bead analysis maintains the relative ordering of the residuals and simply shifts them all by the same amount (wrapping around the start/end of the data), so that correlated noise present in the residuals is preserved. A prayer-bead analysis is not appropriate if the nature of the noise is expected to change throughout the observations (e.g., the 8 $\mu$m ramp seen in cryogenic \emph{Spitzer} observations).  But there is no evidence for such changes in behavior in \emph{Warm Spitzer} data in general, or in our time series in particular.  The prayer-bead can only have as many iterations as there are data points, but this is not a problem for the current study given our approximately 52,000 images; we run the prayer bead for 10,000 iterations with randomly chosen offsets and verify that the uncertainties have converged by comparing to 100- and 1000-iteration prayer-bead MCs.  It is also worth noting that with such a long data set, we are more likely to observe rare instances of bad behavior in the detector, making the prayer-bead technique particularly conservative.  Indeed, prayer-bead Monte Carlo provides the largest error bars for the phase variation parameters: $A_{\rm therm}$, $\alpha_{\rm max}$, $A_{\rm ellips}$.

\section{Results}
The two methods used to remove intra-pixel sensitivity variations are fundamentally different. The Gaussian decorrelation uses local information to correct the flux with no assumption about larger-scale trends; it is able to correct for small-scale variations in sensitivity, but requires very high densities of centroids. The polynomial fit assumes a functional form for the smoothly-varying sensitivity, but is better able to correct regions that are less-well sampled.  It is not clear which method is better suited to a given data set. \cite{Ballard_2010} found the decorrelation to be better (in terms of $\chi^{2}$) than the polynomial fit when analyzing their binned 4.5~$\mu$m time series. We run both methods on unbinned data and find that the two methods perform equally well at 4.5 $\mu$m, while at 3.6 $\mu$m the polynomial fit is better.

Given their very different underlying philosophies, it is encouraging that the two methods yield similar weight functions (compare the pixel maps in Figure 3 to those in Figures 4 and 5).  In fact, the transit depths, eclipse depths, and ellipsoidal variations recovered by the two techniques are generally consistent. The thermal phase variation amplitude and offset differ significantly, however.  

\subsection{Transits}
Our values for the impact parameter and geometrical factor are broadly consistent with published values.  
Note that we simultaneously fit the transits and eclipses, and eclipses are notoriously bad at constraining these orbital parameters.\footnote{If we only consider the 0.3 days of data centered on each transit, we find impact parameters of $b=0.46$ and 0.60 (larger than published values), and geometrical factors of $a/R_{*}=2.9$ and 2.7 (smaller than published values) at 3.6 and 4.5 $\mu$m, respectively.}

Combining our transit times for epochs 925 and 947 (BMJD of 55,518.0407(4) and 55,542.0521(4), respectively) with those of \cite{Chan_2011} and the ephemeris of \cite{Maciejewski_2011}, we obtain a BJD discovery epoch transit center of $t_{0} = 2454508.9768(2)$ and period $P=1.0914207(4)$~days. Since our transits occurred slightly earlier than predicted, our best-fit orbital period is $3.5\sigma$ shorter than that of \cite{Maciejewski_2011}, but it is notoriously difficult to compare transit times at different wavelengths analyzed by different groups because of subtle differences in star-spot coverage, treatment of limb-darkening, etc. \citep[][]{Desert_2011}.

More importantly, we are simultaneously fitting an entire orbit worth of data including many different astrophysical effects, while the optical transit data were fit on their own, irrespective of longer-term astrophysical trends. In order to make a more appropriate comparison, we try fitting the transits independently of the rest of the data (considering only those data within 0.15 days of the transit center). This yields later transit centers, leading to an ephemeris more consistent with that of \cite{Maciejewski_2011}: $t_{0} = 2454508.9767(2)$, $P = 1.0914210(4)$.  

\begin{figure}[htb]
\includegraphics[width=84mm]{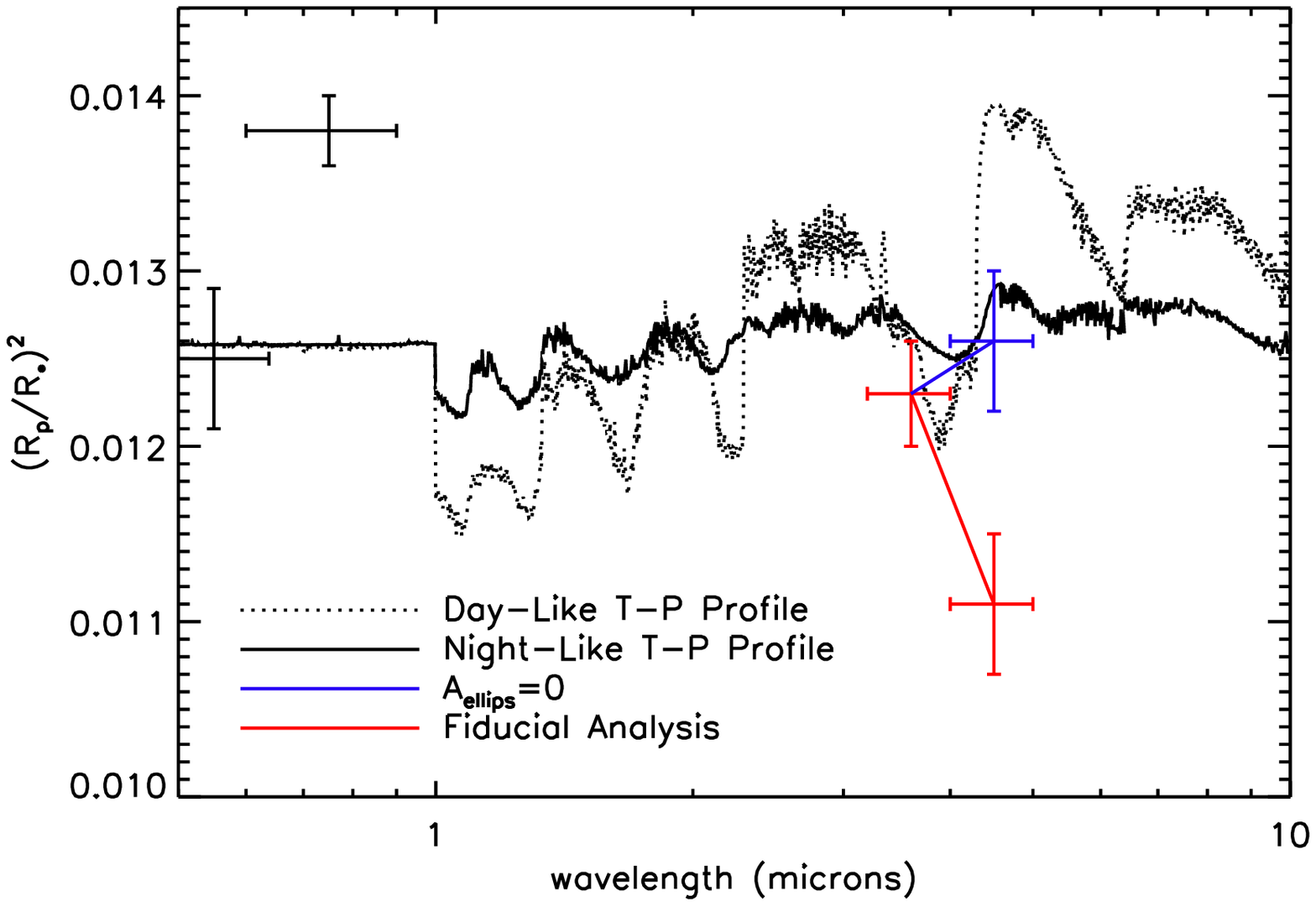}
\caption{The predicted wavelength-dependent transit depth of WASP-12b based on a Solar composition \citep[][]{Burrows_2007, Burrows_2008}. The dotted black line shows a model with a day-like T-P profile (large scale height); the solid line shows a model with a night-like T-P profile (short scale height). The red points correspond to a model with ellipsoidal variations; the blue point corresponds to a model without 4.5 $\mu$m ellipsoidal variations (see first Section~5.3.2). The two black points with error bars on the left show the optical transit depth from \cite{Hebb_2009} and \cite{Maciejewski_2011} (top) and \cite{Chan_2011} (bottom); we normalize the model transit spectrum to the latter's observation.} 
\label{burrows_transit}
\end{figure}

Because of WASP-12b's high temperature and inflated radius, the variations in transit depth with wavelength should be $3\times$ larger than for ``typical'' hot Jupiters \citep[using the scaling relation from][]{Winn_2010}. Based on observations and models of HD~189733b \citep{Fortney_2010}, one might therefore expect the transit depth at 4.5 $\mu$m to be $\sim10^{-3}$ deeper than at 3.6 $\mu$m. This is borne out by the dotted black line in Figure~6, which shows a \cite{Burrows_2007, Burrows_2008} model transit spectrum assuming Solar composition and a day-like temperature-pressure (T-P) profile.  If the planet's terminator has a night-like T-P profile (shorter scale height; solid black line in Figure~\ref{burrows_transit}), the difference in transit depth could be as small as $2\times 10^{-4}$ ---but always with a transit depth greater at 4.5 $\mu$m than at 3.6 $\mu$m.

In Figure~6 we compare our transit depths at 3.6 and 4.5 $\mu$m with previously published optical values: 0.0138(2) \citep[B and z'-band,][]{Hebb_2009}, 0.01380(16) \citep[R-band,][]{Maciejewski_2011}, and 0.0125(4) \citep[V-band,][]{Chan_2011}, yielding a three-band transmission spectrum of the planet. 
Curiously, the transit depth at 4.5 $\mu$m is considerably shallower than at 3.6 $\mu$m. If we adopt the larger \cite{Maciejewski_2011} optical transit depth, then our mid-IR transit depths
 could be indicative of hazes or optical absorbers in the atmosphere of WASP-12b, as has been inferred for HD~189733b \citep{Pont_2008, Sing_2011}.  The larger radius at 3.6 $\mu$m as compared to 4.5 $\mu$m indicates a higher atmospheric opacity at the shorter waveband, which is difficult to reconcile with current models.

In an attempt to explain its peculiar eclipse spectrum, \cite{Deming_2011} hypothesized that CoRoT-2b might have equal opacity at 3.6 and 4.5 $\mu$m (and lower opacity at 8 $\mu$m) due to a haze of ---as yet unknown--- $\mu$m-sized particles.  One may similarly explain the unusual WASP-12b transit spectrum in terms of a haze of slightly smaller particles, such that the opacity drops from 3.6 to 4.5 $\mu$m. Given the large difference in transit depth between the two wavebands, this hypothesis also requires a large atmospheric scale height at the planet's terminator. 

\subsection{Eclipses}
At 3.6 $\mu$m, the eclipse depth is $\sim1\sigma$ lower using the polynomial fit as compared to the decorrelation. If we fit the two eclipses individually using the polynomial IPSV fit, we obtain depths of 0.0030 (highly correlated residuals) and 0.0038 (incomplete egress), respectively. Given the prior measurement of 0.0038(1) by \cite{Campo_2011}, we adopt the larger value from the Gaussian decorrelation for our analysis (this choice does not significantly affect any of our conclusions).

At 4.5 $\mu$m, we obtain comparable $\chi^{2}$ values and eclipse depths regardless of our treatment of systematics. In both cases they are consistent with the \cite{Campo_2011} value: 0.0038(2). 

In all cases our error estimates are somewhat larger than the \cite{Campo_2011} estimates. We observed the same planetary system with the same instrument, so one expects the same eclipse depths and uncertainties.  The only significant change in the detector is that our observations occurred after the fall 2010 change in heater cycling, as mentioned in Section~2.  This means that the short-term telescope jitter for our observations has a period of 30 minutes rather than 1 hour, and the amplitude of the centroid excursions ---and hence flux variations--- is reduced by a factor of two (for reference, the ingress/egress time for WASP-12b is 21 minutes, and the total transit/eclipse duration is 3 hours). 

It is conceivable that the near coincidence between the centroid jitter half-period and the ingress/egress timescale leads to greater residual systematics in our data than in the \cite{Campo_2011} time series.  However, our eclipse depths are based on two occultations, and we have a much longer baseline of observations to help us correct for detector systematics, characterize noise properties and estimate uncertainties (see Section 4.2).  Our MCMC error estimates are similar to those of \cite{Campo_2011}, but residual-binning and prayer-bead analyses produce eclipse depth uncertainties more than $2\times$ larger than the MCMC.  This means that there is still red noise present in our residuals, and the prayer-bead analysis is a more realistic estimate of our parameter uncertainties.

\subsubsection{Day-Side Emergent Spectrum}
If one assumes solar composition, the relative eclipse depths at 3.6 and 4.5 $\mu$m can probe the temperature vs. pressure (T-P) profile of the planet. Water vapor absorbs less at 3.6 than at 4.5 $\mu$m, so the shorter waveband will have a higher brightness temperature if the temperature is locally dropping with height, or vice versa \citep[e.g.,][]{Burrows_2010book}.  

If the eclipse depth is measured at sufficiently many wavelengths, one may hope to simultaneously constrain a planet's atmospheric composition and T-P profile \citep[e.g.,][]{Madhusudhan_2009b}. The high C/O chemistry invoked by \cite{Madhusudhan_2011} was the result of an abnormally low eclipse depth at 4.5 $\mu$m in the \cite{Campo_2011} data, which was interpreted as being due to CO absorption in the planet's relatively cool upper atmosphere. More recently, \cite{Kopparapu_2011} used a photochemical model to study the disequilibrium chemistry of WASP-12b, confirming that CO would be enhanced in a high C/O composition atmosphere.

\begin{figure}[htb]
\includegraphics[width=84mm]{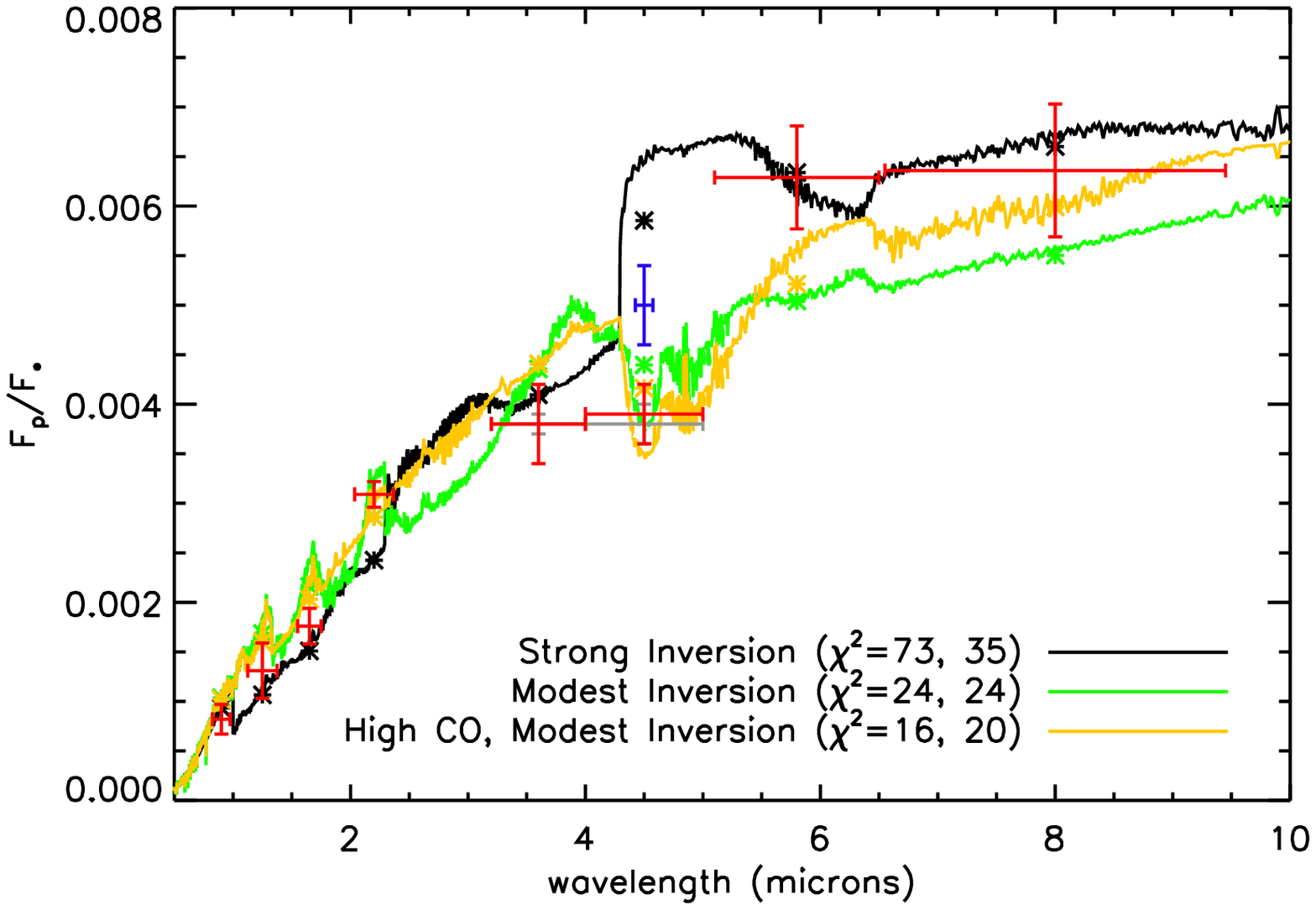}
\caption{The day-side emergent spectrum of WASP-12b.  The colored lines show various 1D atmospheric models \citep[][]{Burrows_2007, Burrows_2008}, while the red points show the measured secondary eclipse depths.  From left to right, the data are: z-band \citep{Lopez-Morales_2010}; J, H, and K$_{s}$-band \citep{Croll_2011}, IRAC channels 1 \& 2 (this study); and IRAC channels 3 \& 4 \citep{Campo_2011}. Note that the K$_{s}$-band eclipse depth and the K-H eclipse color have been confirmed by \cite{Zhao_2011} and \cite{Crossfield_2012}, respectively. The \cite{Campo_2011} eclipse depths at 3.6 and 4.5 $\mu$m are shown in gray. The blue point shows the 4.5 $\mu$m eclipse depth if ellipsoidal variations are set to zero, the ``null hypothesis''. In the legend, the first $\chi^{2}$ value for each model is for the fiducial analysis (including ellipsoidal variations), the second value is for the null hypothesis (setting ellipsoidal variations to zero). The enhanced-CO model (yellow line) offers the best fit, but the Solar composition model (green line) is not significantly worse.}
\label{burrows}
\end{figure}

In Figure~\ref{burrows}, we compare the near- to mid-IR broadband spectrum of WASP-12b to various 1-D radiative transfer models \citep[][]{Burrows_2007, Burrows_2008}. We vary the abundance of CO as a proxy for varying the C/O ratio. Our eclipse depths are consistent with those of \cite{Campo_2011}, so we still favor models with enhanced CO (10$\times$ Solar) and a weak inversion for this planet, in agreement with \cite{Madhusudhan_2011}. We also find that we can obtain an equally good fit to the data by reducing H$_{2}$O to 1\% Solar abundance and partitioning carbon evenly between CO and CH$_{4}$. Given the caveat that these models are in radiative ---but not chemical--- equilibrium, the crux of fitting WASP-12b's unique day-side spectrum is to suppress H$_{2}$O with respect to CO.  

Our larger error bars, however, make these composition statements marginal: the Solar composition model with modest inversion (the green line in Figure~\ref{burrows}) is only worse by $\Delta\chi^{2}=8$ for 8 degrees of freedom.  Furthermore, neither the standard composition nor the enhanced CO scenario are consistent with the relative transit depths at 3.6 and 4.5 $\mu$m (s.f. Figure~\ref{burrows_transit} and previous section). It may be possible to reconcile these two measurements if the atmospheric composition is grossly different at the day-night terminator than near the sub-stellar point.  

\subsection{Ellipsoidal Variations}
The best-fit ellipsoidal variations at 3.6 $\mu$m are consistent with the predicted amplitude of $2\times10^{-4}$, but as one can see from the relative uncertainty (or from glancing at Figure~\ref{wasp_12_slope_ch1}), they are not robustly detected: the $\chi^{2}$, residuals and remaining astrophysical parameters do not change significantly if $A_{\rm ellips}$ is set to zero.

As shown in Figure~5, however, we clearly detect power at 4.5 $\mu$m in the second cosine harmonic, $\cos(2\alpha)$, consistent with the prediction of ellipsoidal variations due to the prolate shape of the planet \citep{Li_2010, Leconte_2011, Budaj_2011}. The semi-amplitude of the variations, however, is $1.2(2)\times10^{-3}$ using either decorrelation or polynomial IPSV-removal, approximately $6\times$ the predicted value. 

Since ellipsoidal variations are primarily a geometrical effect, it is difficult to understand how the measured amplitude could be be so different at 3.6 and 4.5 $\mu$m. The upper layers of the atmosphere should be more distorted; the deeper layers more spherical. 
The relative strengths of ellipsoidal variations at the two wavebands imply that the 4.5 $\mu$m flux is originating from much higher up in the planet's atmosphere than the 3.6 $\mu$m flux.  The simplest way to do this is for the atmosphere to have a greater opacity at 4.5 $\mu$m than 3.6 $\mu$m.  
But the relative transit depths indicate exactly the opposite, as described above and shown in Figure~\ref{burrows_transit}.  

Alternatively, it is possible that detector systematics still present after IPSV-removal attenuate the ellipsoidal signal at 3.6 $\mu$m or enhance it at 4.5 $\mu$m. The raw photometry shown in Figure~1 implies that the 4.5 $\mu$m light curve is more trustworthy of the two. We therefore begin by assuming that the ellipsoidal signal at 4.5 $\mu$m is entirely astrophysical in nature, then consider the opposite scenario.

\subsubsection{Interpreting the Ellipsoidal Variations at 4.5~$\mu$m}
If we take the 4.5 $\mu$m ellipsoidal variations at face value, they have surprising implications for the planet's shape. The dimensions of a prolate planet may be described by its short and long radii, denoted by $R_{p}$ and $R_{\rm long}$, respectively. The third dimension (parallel to the system's angular momentum vector) is assumed to be $R_{p}$ because we are neglecting rotational effects, which tend to produce oblate planets. Note that WASP-12b is on a very short-period orbit, so ---if tidally locked--- it has a rotation rate only a factor of two slower than Jupiter or Saturn. As a result of this rotation, \cite{Budaj_2011} estimates the planet's polar radius to be 2.5\% shorter than its lateral equatorial radius,\footnote{Note, furthermore, that the sub-stellar and anti-stellar planetary radii are not equal.  Nevertheless, the dominant effect is the planet's prolate shape.} so strictly speaking WASP-12b is a triaxial ellipsoid, but this does not affect the analysis below because of the planet's edge-on orbit and our relatively short baseline of observations is insensitive to the spin precession of the planet \citep{Carter_2010b}.   

At conjunction ---either transit or eclipse--- we are seeing the planet's smallest projected area ($\pi R_{p}^{2}$, in the case of a perfectly edge-on orbit).  The projected area of an ellipsoid on an edge-on orbit varies as \citep[e.g.,][]{Vickers_1996}:
\begin{equation}
A_{p}(\alpha) = \pi R_{p}^{2} \sqrt{\cos^{2}\alpha + \left(\frac{R_{\rm long}}{R_{p}}\right)^{2}\sin^{2}\alpha}. 
\end{equation}
For $R_{\rm long}/R_{p} \approx 1$, the changes in projected area follow a $A_{p} \propto \cos(2\alpha)$ shape and this is in fact how we modeled them; for more severe elongations, the peaks become broader and the troughs narrower, until a limiting case of $A_{p} \propto |\sin\alpha|$. 

\begin{figure}[htb]
\includegraphics[width=84mm]{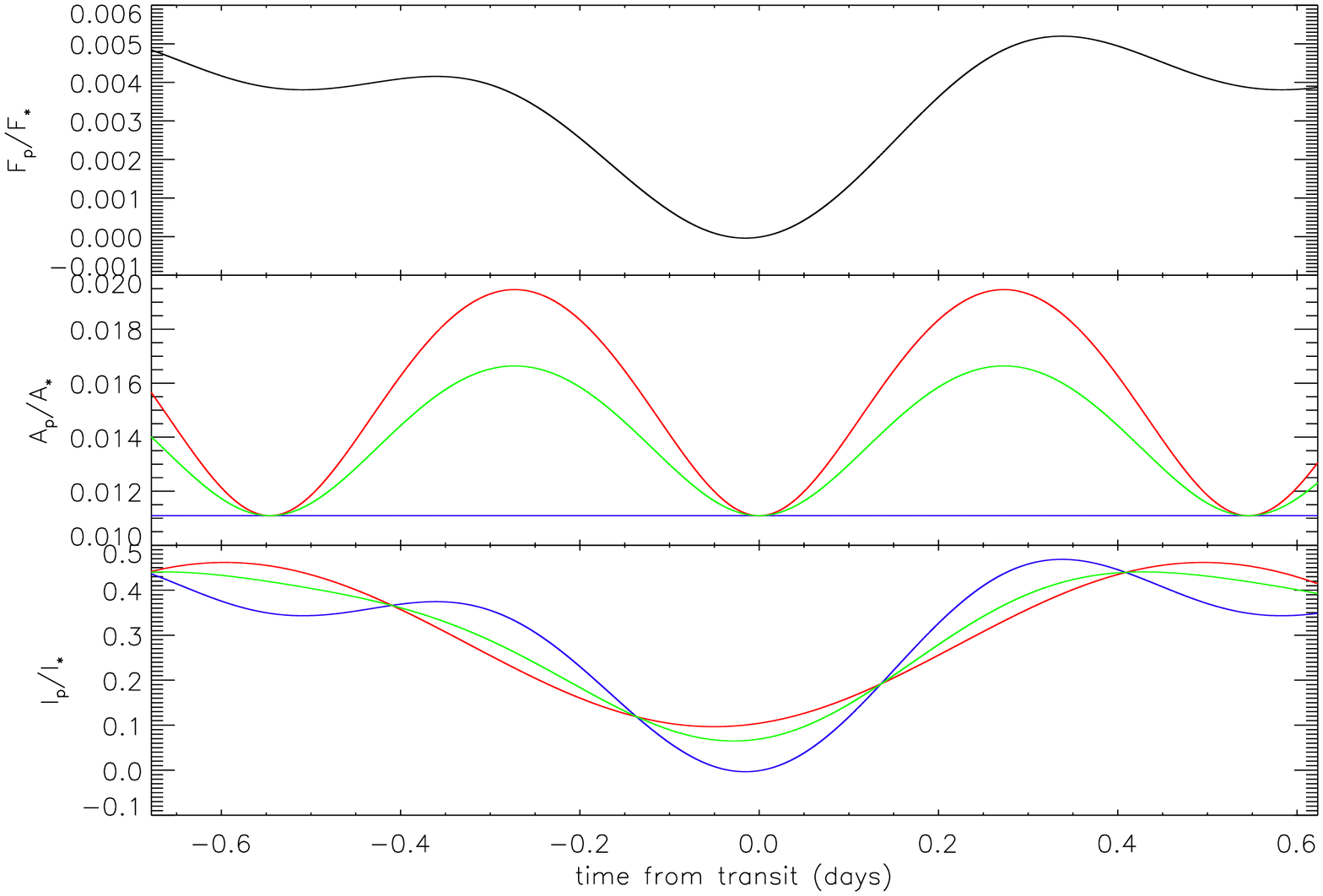}
\caption{Our best-fit 4.5 $\mu$m phase variations (top panel) can be modeled as a geometrical component due to the planet's changing projected area (middle panel) and a thermal component due to longitudinal variations in the planet's brightness (bottom panel). For the red lines all of the $2\alpha$ power is attributed to the planet's ellipsoidal shape; this hypothesis can be ruled out by the transit morphology (Figure~9).  The blue lines result from assuming that the planet is spherical in shape; this hypothesis leads to unphysical brightness variations on the planet.  The green lines show the middle-road: most of the $2\alpha$ power is due to the planet's shape, but the thermal component also contributes a bit. Note that this \emph{a posteriori} analysis has limitations: in our astrophysical model, the thermal and ellipsoidal components of phase variations were added, while strictly speaking the planet's intensity and cross-sectional area should be multiplied.}
\label{ellipsoidal}
\end{figure}

If we interpret all of the power in the second cosine harmonic as being due to the changing cross-sectional area of the planet (the red curves in Figure~\ref{ellipsoidal}), we may estimate the planet's aspect ratio as 
\begin{equation}
R_{\rm long}/R_{p} \approx 1 + 2A_{\rm ellips}\langle F_{*}/F_{\rm day}\rangle = 1.8(1). 
\end{equation}
The predicted aspect ratio for the planet is $R_{\rm long}/R_{p}=1.1$, while the aspect ratio for a Roche lobe is $3/2 = 1.5$.  

\begin{figure}[htb]
\includegraphics[width=84mm]{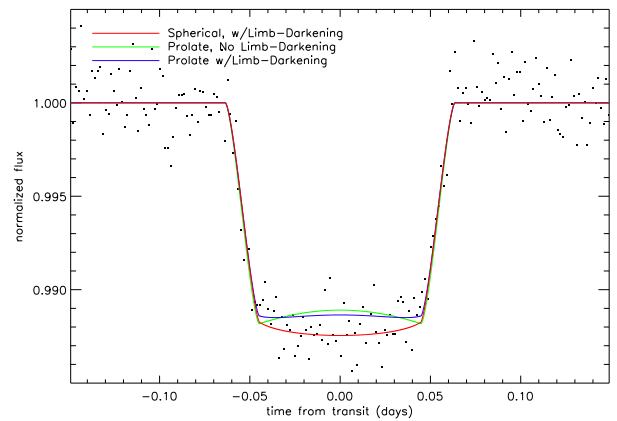}
\caption{WASP-12b transit lightcurve at 4.5 $\mu$m; the data have been binned for plotting. The red line shows our fiducial model: a spherical planet with non-linear limb-darkening of the star. The green line shows the effect of the planet's changing cross-sectional area (but not its changing shape). The blue line shows how limb-darkening partially washes out this signal.  For the two prolate planet models, we use $R_{\rm long}/R_{p}=1.8$, the most extreme scenario supported by our phase variations.}
\label{transit_morphology}
\end{figure}

It is worth noting that a planet with an aspect ratio of $1.8$ would have a 13\% larger projected area at the start and end of transit as compared to transit center, resulting in a w-shaped transit, in the absence of stellar limb-darkening \citep[note that this differs from changes in transit morphology due to an oblate planet discussed in][]{Carter_2010}.  Since the transit depth of WASP-12b is approximately 1\%, this shape-induced transit effect would come in at the $1.3\times10^{-3}$ level and might be detectable in our current data.  In Figure~\ref{transit_morphology} we estimate the expected transit morphology by treating the planet as a sphere with variable cross-sectional area. The red line shows our fiducial model: a spherical planet with non-linear limb-darkening of the star.  The green line shows the effect of the planet's changing cross-sectional area (but not its changing shape). The blue line shows how limb-darkening partially washes out this signal.  Figure~\ref{transit_morphology} implies that our data rule out the most extreme prolate toy model, but clearly the treatment of limb-darkening is important here.  

If we instead interpret the phase curve as being caused entirely by longitudinal brightness variations on a spherical planet (the blue curves in Figure~\ref{ellipsoidal}), it can be inverted into a longitudinal intensity map of the planet, following \cite{Cowan_2008}.  Because of the low-pass filtering that occurs in the map $\to$ lightcurve convolution, the deconvolution will enhance the highest frequency terms present in the lightcurve \citep[e.g., Figure~2 of ][]{Cowan_2009proc}.  In the current case, the resulting map has two prominent temperature peaks: one at the dawn terminator and one at the dusk terminator. More importantly, the only way to simultaneously fit the bright terminators and dark night-side is by having negative intensity at the anti-stellar point, clearly an unphysical solution.   

Another argument against a spherical planet is that while the thermal phase variations of a spherical planet will in general contain power in the second harmonic, there is no reason for it to all appear in the cosine rather than sine term.\footnote{This is in stark contrast to the first harmonic, where one expects most of the power to be in the $\cos\alpha$ term due to the extreme day-night forcing.} When we run fits allowing for the phase of the $\cos(2\alpha)$ term to vary, the offset is consistent with zero at the 2$\sigma$ level, with the largest offset being less than $6^{\circ}$. This indicates that the flux is peaking at quadrature, as expected for ellipsoidal variations. To our knowledge, there is no reason to expect such a temperature profile if the planet is spherical. 

If the planet is prolate, however, one might expect something akin to the gravity darkening/brightening seen in some binary stars: the sub-stellar and anti-stellar points on the planet are farther from the center of the planet than the terminator, leading to lower surface gravity.  The lower surface gravity at the sub-stellar and anti-stellar  regions might lead to cooler temperatures, all things being equal \citep{VonZeipel_1924}. But this would only affect the intrinsic component of the planet's power budget, expected to be insignificant for hot Jupiters.  

Finally, it is possible that the planet's shape affects the circulation of its atmosphere in more subtle ways, leading to relatively hot regions at the dawn and dusk terminators. This brings us to our favored astrophysical interpretation of the ellipsoidal variations. The planet may have an aspect rati oof $R_{\rm long}/R_{p} \approx 1.5$ (somewhat easing the transit morphology constraints described above), with an additional enhancement of the $\cos(2\alpha)$ power because of a relatively hot day-night terminator (the green curve in Figure~\ref{ellipsoidal}). Since the atmospheric dynamics on severely non-spherical planets has not yet been addressed in the literature, it is difficult to say whether this scenario is reasonable.
  
\subsubsection{The Null Hypothesis at 4.5~$\mu$m}
If the amplitude of ellipsoidal variations, $A_{\rm ellips}$, is set to zero at 4.5 $\mu$m, the residuals become more correlated (as one would guess from Figure~5), increasing the red noise and hence uncertainties in the other astrophysical parameters. Furthermore, including terms up to sixth-order in the polynomial IPSV-removal does not obviate the need for the ellipsoidal term, and the same signal is detected in the Gaussian decorrelation version of the analysis. The $\chi^{2}$ is worse when ellipsoidal variations are ignored ($\Delta \chi_{R}^{2} =0.007$ for either the decorrelation or polynomial fit).  

However, it is conceivable that the 4.5 $\mu$m ellipsoidal signal is in fact uncorrected detector systematics.  It is therefore worth briefly considering the astrophysical implications of this scenario.

The most obvious implication of the null hypothesis is that the Roche-filling upper-atmosphere of the planet need not be optically thick.  Since the predicted amplitude of ellipsoidal variations were only at the $2\sigma$ level, the null hypothesis is consistent with the predictions of \cite{Li_2010}, \cite{Leconte_2011} and \cite{Budaj_2011}. 

Furthermore, the ellipsoidal variations have minima at inferior and superior conjunction, so the null hypothesis causes the transit depth and eclipse depths to significantly increase. Notably, the 4.5 $\mu$m transit depth becomes 0.0126(4), comensurate with that at 3.6 $\mu$m. The null hypothesis transit depths are consistent with a short (night-like) scale height and Solar composition (solid black line in Figure~\ref{burrows_transit}). 

The deeper 4.5 $\mu$m eclipse depth, $F_{\rm day}/F_{*}=$ 0.0050(4), no longer favors the enhanced CO scenario (yellow line in Figure~\ref{burrows_transit}) invoked by \cite{Madhusudhan_2011}: the fit to the Solar composition model with modest inversion (green line in Figure~\ref{burrows}) is only worse by $\Delta\chi^{2} \approx 4$, for 8 degrees of freedom.

\subsection{Thermal Phase Variations}
We model thermal phases using only first harmonics ($\cos\alpha$ and $\sin\alpha$), while our parameterization of ellipsoidal variations is a second harmonic ($\cos2\alpha$).  These functions are by definition orthogonal, so it is not surprising that our conclusions about thermal phase variations ($A_{\rm therm}$, $\alpha_{\rm max}$) described below are not significantly affected by the presence or absence of ellipsoidal variations discussed above.

That said, the amplitude and offset we obtain for the thermal phase variations depends on which IPSV-removal scheme we use. Unfortunately, it is not clear how to perform a direct model comparison between decorrelation and polynomial fits using the BIC. 
The Gaussian decorrelation can be thought of as having a large number of free parameters and a somewhat smaller number of additional constraining equations, but it is not obvious how to estimate its degrees of freedom. As discussed by \cite{Ballard_2010}, the point-by-point de-correlation does a great job of removing the short-term jitter and long-term detector drift, but may also remove any longer-term astrophysical signal. (This did not interfere with their goal of searching for transits.) Insofar as diurnal phase variations are the most gradual astrophysical signal in our study, it is not surprising that it is the most dependent on IPSV-removal.

The polynomial fit leads to $\chi^{2}$ values at least as good ---and sometimes significantly better--- than the Gaussian decorrelation.  More importantly, the scatter in the residuals on the critical 21 minute timescale is lower for the polynomial fits, indicating that this method is doing a better job of removing correlated noise. 

Our 4.5 $\mu$m pixel sensitivity map obtained from Gaussian decorrelation (top right panel of Figure~3) shows a valley at $y\approx 14.75$; this doesn't follow the usual pattern of sensitivity decreasing towards pixel edges. (Note that we observe the same ripples in sensitivity as \cite{Ballard_2010}, but those occur on a much smaller spatial scale ---and exhibit a much smaller amplitude--- than the $y\approx 14.75$ valley.)  

Finally, the middle-left panel of Figure~3 suggests that the decorrelation method has over-corrected the 3.6 $\mu$m system flux near superior-conjunction: the eclipse bottoms ---which should be flat since the planet is hidden from view--- slope upward towards the central transit.   

We therefore argue that the decorrelation has filtered out much of the phase variations along with the systematics and adopt the polynomial-fitted thermal phase parameters in what follows.\footnote{For completeness, we include the thermal phase parameters resulting from the decorrelation in Tables 2 and 3. That analysis leads to smaller phase amplitudes than the polynomial fit. If taken at face value, this implies lower albedo and higher heat transport efficiency.}
 
\begin{figure}[htb]
\includegraphics[width=84mm]{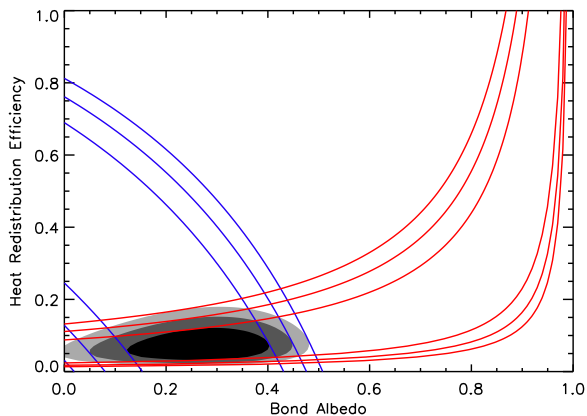}
\caption{1, 2, and $3\sigma$ constraints on the Bond albedo and recirculation efficiency of WASP-12b from thermal eclipse (blue) and phase variation (red) observations, using the parametrization of \cite{Cowan_2011b}.  The gray-scale shows the confidence intervals for the combined constraints.}
\label{albedo_epsilon}
\end{figure}

Following \cite{Cowan_2011b}, we estimate the hemispheric effective temperatures to be  $T_{\rm day}=2928(97)$~K and $T_{\rm night}=983(201)$~K, where the uncertainties include an estimate of systematic errors in going from brightness temperatures to effective temperatures.\footnote{WASP-12b has a sub-stellar equilibrium temperature of $T_{0} = 3555(132)$~K, a no-albedo, no-recirculation day-side temperature of $T_{\varepsilon=0}=(2/3)^{1/4}T_{0}=3213(119)$~K, and a no-albedo full-recirculation global temperature of $T_{\rm uni}=(1/4)^{1/4}T_{0}=2514(92)$~K \citep[e.g.,][]{Cowan_2011b}.} This implies a Bond albedo of $A_{B}=0.25$ (presumably due to Rayleigh scattering and/or reflective clouds) and very low heat recirculation efficiency, $\varepsilon<0.1$, at 1$\sigma$ (see Figure~\ref{albedo_epsilon}).\footnote{If one presumes that advection is the dominant mode of heat transport, then $\varepsilon \approx \tau_{\rm rad}/(\tau_{\rm adv}+\tau_{\rm rad})$, where $\tau_{\rm rad}$ and $\tau_{\rm adv}$ are the characteristic radiative and advective timescales at the mid-IR photosphere.}  Note that our 1D radiative transfer models used for interpreting transit and eclipse spectra are cloud-free and have an albedo lower than this inferred value.

Space-based optical secondary eclipse depths have been measured for a handful of hot Jupiters.  If the planet's equilibrium temperature is sufficiently low, such measurements provide an unambiguous estimate of geometric albedo, $A_{g}$: $0.04(5)$ for HD~209458b \citep{Rowe_2008}, $0.32(3)$ for Kepler-7b \citep{Demory_2011}, $0.10(2)$ for Kepler-17b \citep{Desert_2011b}, $0.30(8)$ for KOI-196b \citep{Saterne_2011}, and $0.025(7)$ for TrES-2b \citep{Kipping_2011}.  Acknowledging that WASP-12b is more than 1000~K hotter than any of those planets, a Bond albedo of 0.25 is well within the observed range for hot Jupiters. 

Assuming gray albedo and a Lambertian scattering phase function \citep[$A_{B}=\frac{3}{2}A_{g}$;][]{Hanel_1992}, the reflected-light secondary eclipse of WASP-12b should have a depth of $2.4\times10^{-4}$, comparable to current ground-based precision for this target. In practice, most planets exhibit an opposition surge (making them disproportionately bright at superior conjunction) and Rayleigh scattering, so the actual contrast ratio in blue optical wavebands is likely more favorable than this estimate. 

Assuming Solar atmospheric composition and hence opacity, the 3.6 $\mu$m thermal flux should originate from deeper in the atmosphere than any other mid-IR waveband. Insofar as radiative times increase monotonically with pressure, we may therefore expect the 
3.6 $\mu$m phase variations to be muted compared to the 4.5 $\mu$m phase variations, and the phase offset should be greater at the shorter wavelength \citep[e.g., Figure 9 of][]{Burrows_2010}.

The hot-spot offset is $53(7)^{\circ}$ East of the sub-stellar point at 3.6 $\mu$m. While not as extreme as $\upsilon$-Andromeda~b \citep{Crossfield_2010}, it is difficult to reconcile our large phase offset and large amplitude at 3.6 $\mu$m.\footnote{It is tempting to attribute the early peak of the 3.6 $\mu$m phase curve to Doppler beaming, which would produce a peak in flux when the star is moving towards us, a quarter period before superior conjunction \citep{Loeb_2003}.  The expected amplitude of this signal, however, is only $7.5\times10^{-7}$ on the Rayleigh-Jeans tail of the star.} It is worth noting, however, that there are highly-correlated residuals near the purported peak ($\sim0.35$~day after transit) which may be partially responsible for the large offset.

On the other hand, we find that the 4.5 $\mu$m phase amplitude and offset, $16(4)^{\circ}$ E, are consistent with a \cite{Cowan_2011a} model with heat transport efficiency of $\epsilon\equiv\tau_{\rm rad}/\tau_{\rm adv}\approx0.1$. To put this in context, phase variations and eclipse timing offset of HD~189733b at 8 $\mu$m indicate $\epsilon \approx 0.7$ \citep{Agol_2010}.
  
\subsubsection{Implications of Thermal Phase Variations}
\cite{Cowan_2011b} noted that the day-side temperatures of the hottest short-period giant planets are very close to the theoretical upper limit of no albedo and no recirculation.  This was in contrast to run-of-the-mill hot Jupiters (e.g., HD~189733b, HD~209458b), which exhibit a variety of albedos/recirculation efficiencies, albeit consistent with generally low albedos ($A_{B}<0.3$). In a statistical study of \emph{Kepler} planetary candidates, \cite{Coughlin_2011} also found generally low albedos for hot Jupiters based on optical secondary eclipses.

The amplitude of the phase variations for WASP-12b depends on the details of the systematics correction, but ---for reasons stated at the start of Section~5.4--- we favor the polynomial fit, which implies a large day--night temperature contrast, and a non-zero Bond albedo.  This suggests that the difference between the hottest short-period giant planets and other hot Jupiters is not albedo, but recirculation efficiency.  (Differences in albedo may very well explain the differences in day-side effective temperature amongst the remaining hot Jupiters, however.) 

What could make the hottest of hot Jupiters poor heat re-circulators? There are two classes of solutions: decreasing either the planet's characteristic advective frequency, or radiative timescale.\footnote{One can imagine more exotic means of transporting energy \citep[e.g., gravity waves;][]{Watkins_2010} but most hydrodynamical simulations suggest that horizontal energy transport on hot Jupiters is primarily a matter of advection.} 

\emph{More Magnetic Drag:} Assuming these planets have magnetic fields, the movement of ionized alkali metals through the field produces drag that is collisionaly imparted on the dominant neutral species (presumably H and He). Hotter planets should have more ionized species, more drag and therefore a harder time advecting heat to their night side \citep{Perna_2010}.  Because of the non-linear dependence of ionization on temperature, this effect could lead to sudden changes in dynamical regime as one considers increasingly hot planets. The scaling relations of \cite{Menou_2011} indicate that the temperature above which magnetic drag severely curtails heat transport is inversely related to the planet's magnetic field strength.  Even the weakest field they considered in their study, 3~Gauss, would result in very low recirculation efficiency for a planet as hot as WASP-12b. 

\emph{Shorter Radiative Times:} Following the argument of \cite{Cowan_2011b}, the radiative relaxation time of a parcel of gas scales as $\tau_{\rm rad}\propto T^{-3}$ \citep[][]{Iro_2005, Seager_2005}. But zonal wind speeds may also increase with the amplitude of the diurnal forcing. If one assumes that the wind speeds have a fixed Mach number, they should scale as $v_{\rm wind}\propto T^{1/2}$, and therefore the advective time should scale as $\tau_{\rm adv}\propto T^{-1/2}$ \citep[a more detailed scaling analysis leads to the same temperature dependence;][]{Menou_2011}. The stronger dependence on temperature of radiative time compared to advective time implies that ---all things being equal--- hotter planets should be less efficient at balancing their day--night temperature contrast. This effect should cause the heat transport efficiency to gradually decrease as one considers increasingly hot planets.

\emph{Weaker Greenhouse:} Atmospheric opacity is typically greater at the thermal wavelengths of emergent radiation than at the visible wavelengths of incident radiation. For the hottest planets, the blackbody peak of thermal emission approaches the peak of their host star, so the opacities of the incoming and outgoing streams converge. In that limit, one expects the thermal photosphere and the optical deposition depth to be one and the same: thermal radiation should escape the atmosphere just as easily as the incoming stellar radiation came in. As with the scalings above, this effect should lead to gradually decreasing recirculation efficiency as a function of increasing planet temperature. 

More observations of thermal eclipses and phase variations for hot Jupiters ---especially those near the $T_{0} \approx 2700$~K transition--- will be necessary to distinguish between the magnetohydrodynamic and the radiative timescale arguments.

\section{Conclusions}
We obtained \emph{Warm Spitzer} full-orbit phase observations of WASP-12b at 3.6 and 4.5 $\mu$m, allowing us to measure the transit depths, eclipse depths, thermal and ellipsoidal phase variations at both wavelengths. We are able to push \emph{Warm Spitzer} photometry to within 10--20\% of the Poisson limit, but there are two important caveats: 

A) Removing intra-pixel sensitivity variations (IPSVs) from the data is inherently a model-dependent endeavor. This means that we must specify not only an IPSV model, but also an astrophysical model before getting close to the quoted precision. The simultaneous fit to astrophysical and systematic effects makes it difficult to produce a ``clean'' lightcurve independent of astrophysical assumptions.  For example, we obtain very different thermal phase variation parameters depending on how we correct for systematics, and it is difficult to distinguish between these scenarios based solely on goodness-of-fit. Instead, we must resort to a number indirect clues as to which IPSV-removal scheme is more trustworthy. 

B) There is still red noise in our residuals, no matter how we remove IPSVs.  This remaining red noise is the dominant source of uncertainty for all of our astrophysical parameters.

We find that WASP-12b exhibits large-amplitude thermal phases ---indicative of poor day--night heat transport and a moderate Bond albedo--- but also an unexpectedly large phase offset at 3.6 $\mu$m.  We do not detect ellipsoidal variations at 3.6 $\mu$m, while we detect an unexpectedly strong signal at 4.5 $\mu$m.  This leads us to two possible hypotheses:

1) If we take the 4.5~$\mu$m ellipsoidal variations at face value, we find: deeper transits at 3.6 $\mu$m as compared to 4.5 $\mu$m, inconsistent with either Solar or enhanced CO models; eclipse depths consistent with previous studies. If the 4.5 $\mu$m ellipsoidal variations are astrophysical in nature, it indicates that the planet is far more distorted than predicted, and exhibits a bright terminator. In this scenario, the 3.6 $\mu$m ellipsoidal variations are attenuated due to detector systematics, possibly throwing off the 3.6 $\mu$m transit depth as well. 

2) If instead we presume that the 4.5 $\mu$m ellipsoidal variations are caused by detector systematics and set them to zero ---the null hypothesis--- we find: transit depths consistent with a Solar composition and short atmospheric scale height at the planet's terminator; eclipse depths consistent with a Solar composition and a modest temperature inversion; ellipsoidal variations in line with predictions.

The null hypothesis is attractive in its simplicity, but requires that we were very unlucky; follow-up \emph{Warm Spitzer} observations would have different systematics (the PSF would fall on different regions of the pixels) and could settle the question of ellipsoidal variations.  It is likely that near infrared transit spectroscopy could break the composition degeneracy, or at least determine the atmospheric structure of WASP-12b; if the planet has a short scale-height at the terminator it will lend credence to the null hypothesis.  Further optical transit photometry will be useful in pinning down the transmission spectrum and refining geometrical parameters; if $a/R_{*}<3$, then the planet could very well be more distorted than predicted, making the large ellipsoidal variations more plausible.  Optical eclipse measurements from the ground or from space might confirm the moderate albedo of the planet.

The planet is hypothesized to be losing mass to its host star.  If this is indeed the case, the presence of an accretion disk, accretion stream and impact hot spot may necessitate a more holistic model to properly interpret observations.  

\acknowledgements
Much of this work was completed while NBC was at the Aspen Center for Physics. NBC acknowledges useful conversations with F.A.~Rasio, W.M.~Farr, H.A~Knutson, N. Lewis and J.~Budaj, as well as countless fruitful discussions at the Future of Astronomy, Extreme Solar Systems II, and joint EPSC/DPS meetings. This work is based on observations made with the Spitzer Space Telescope, which is operated by the Jet Propulsion Laboratory, California Institute of Technology under a contract with NASA. Support for this work was provided by NASA through an award issued by JPL/Caltech.


\begin{thebibliography}{77}
\expandafter\ifx\csname natexlab\endcsname\relax\def\natexlab#1{#1}\fi

\bibitem[{{Agol} {et~al.}(2010){Agol}, {Cowan}, {Knutson}, {Deming}, {Steffen},
  {Henry}, \& {Charbonneau}}]{Agol_2010}
{Agol}, E., {Cowan}, N.~B., {Knutson}, H.~A., {Deming}, D., {Steffen}, J.~H.,
  {Henry}, G.~W., \& {Charbonneau}, D. 2010, \apj, 721, 1861

\bibitem[{{Ballard} {et~al.}(2010){Ballard}, {Charbonneau}, {Deming},
  {Knutson}, {Christiansen}, {Holman}, {Fabrycky}, {Seager}, \&
  {A'Hearn}}]{Ballard_2010}
{Ballard}, S., {Charbonneau}, D., {Deming}, D., {Knutson}, H.~A.,
  {Christiansen}, J.~L., {Holman}, M.~J., {Fabrycky}, D., {Seager}, S., \&
  {A'Hearn}, M.~F. 2010, \pasp, 122, 1341

\bibitem[Budaj(2011)]{Budaj_2011} Budaj, J.\ 2011, \aj, 141, 59

\bibitem[{{Burrows} {et~al.}(2008){Burrows}, {Budaj}, \&
  {Hubeny}}]{Burrows_2008}
{Burrows}, A., {Budaj}, J., \& {Hubeny}, I. 2008, \apj, 678, 1436

\bibitem[{{Burrows} {et~al.}(2007){Burrows}, {Hubeny}, {Budaj}, {Knutson}, \&
  {Charbonneau}}]{Burrows_2007}
{Burrows}, A., {Hubeny}, I., {Budaj}, J., {Knutson}, H.~A., \& {Charbonneau},
  D. 2007, \apjl, 668, L171

\bibitem[{{Burrows} \& {Orton}(2010)}]{Burrows_2010book}
{Burrows}, A. \& {Orton}, G. 2010, in Exoplanets, ed. {Seager, S.}, 419--440

\bibitem[{{Burrows} {et~al.}(2010){Burrows}, {Rauscher}, {Spiegel}, \&
  {Menou}}]{Burrows_2010}
{Burrows}, A., {Rauscher}, E., {Spiegel}, D.~S., \& {Menou}, K. 2010, \apj,
  719, 341

\bibitem[{{Campo} {et~al.}(2011){Campo}, {Harrington}, {Hardy}, {Stevenson},
  {Nymeyer}, {Ragozzine}, {Lust}, {Anderson}, {Collier-Cameron}, {Blecic},
  {Britt}, {Bowman}, {Wheatley}, {Loredo}, {Deming}, {Hebb}, {Hellier},
  {Maxted}, {Pollaco}, \& {West}}]{Campo_2011}
{Campo}, C.~J., {Harrington}, J., {Hardy}, R.~A., {Stevenson}, K.~B.,
  {Nymeyer}, S., {Ragozzine}, D., {Lust}, N.~B., {Anderson}, D.~R.,
  {Collier-Cameron}, A., {Blecic}, J., {Britt}, C.~B.~T., {Bowman}, W.~C.,
  {Wheatley}, P.~J., {Loredo}, T.~J., {Deming}, D., {Hebb}, L., {Hellier}, C.,
  {Maxted}, P.~F.~L., {Pollaco}, D., \& {West}, R.~G. 2011, \apj, 727, 125

\bibitem[{{Carter} \& {Winn}(2009)}]{Carter_2009}
{Carter}, J.~A. \& {Winn}, J.~N. 2009, \apj, 704, 51

\bibitem[{{Carter} \& {Winn}(2010{\natexlab{a}})}]{Carter_2010}
---. 2010{\natexlab{a}}, \apj, 709, 1219

\bibitem[{{Carter} \& {Winn}(2010{\natexlab{b}})}]{Carter_2010b}
---. 2010{\natexlab{b}}, \apj, 716, 850

\bibitem[{{Chan} {et~al.}(2011){Chan}, {Ingemyr}, {Winn}, {Holman},
  {Sanchis-Ojeda}, {Esquerdo}, \& {Everett}}]{Chan_2011}
{Chan}, T., {Ingemyr}, M., {Winn}, J.~N., {Holman}, M.~J., {Sanchis-Ojeda}, R.,
  {Esquerdo}, G., \& {Everett}, M. 2011, \aj, 141, 179

\bibitem[{{Charbonneau} {et~al.}(2005){Charbonneau}, {Allen}, {Megeath},
  {Torres}, {Alonso}, {Brown}, {Gilliland}, {Latham}, {Mandushev}, {O'Donovan},
  \& {Sozzetti}}]{Charbonneau_2005}
{Charbonneau}, D., {Allen}, L.~E., {Megeath}, S.~T., {Torres}, G., {Alonso},
  R., {Brown}, T.~M., {Gilliland}, R.~L., {Latham}, D.~W., {Mandushev}, G.,
  {O'Donovan}, F.~T., \& {Sozzetti}, A. 2005, \apj, 626, 523

\bibitem[{{Claret}(2000)}]{Claret_2000}
{Claret}, A. 2000, \aap, 363, 1081

\bibitem[Coughlin 
\& Lopez-Morales(2011)]{Coughlin_2011} Coughlin, J.~L., \& Lopez-Morales, M.\ 2011, arXiv:1112.1021 

\bibitem[{{Cowan} \& {Agol}(2008)}]{Cowan_2008}
{Cowan}, N.~B. \& {Agol}, E. 2008, \apjl, 678, L129

\bibitem[{{Cowan} \& {Agol}(2009)}]{Cowan_2009proc}
{Cowan}, N.~B. \& {Agol}, E. 2009, in IAU Symposium, Vol. 253, IAU Symposium,
  544--547

\bibitem[{{Cowan} \& {Agol}(2011{\natexlab{a}})}]{Cowan_2011a}
---. 2011{\natexlab{a}}, \apj, 726, 82

\bibitem[{{Cowan} \& {Agol}(2011{\natexlab{b}})}]{Cowan_2011b}
---. 2011{\natexlab{b}}, \apj, 729, 54

\bibitem[{{Cowan} {et~al.}(2007){Cowan}, {Agol}, \& {Charbonneau}}]{Cowan_2007}
{Cowan}, N.~B., {Agol}, E., \& {Charbonneau}, D. 2007, \mnras, 379, 641

\bibitem[{{Croll} {et~al.}(2011){Croll}, {Lafreniere}, {Albert},
  {Jayawardhana}, {Fortney}, \& {Murray}}]{Croll_2011}
{Croll}, B., {Lafreniere}, D., {Albert}, L., {Jayawardhana}, R., {Fortney},
  J.~J., \& {Murray}, N. 2011, \aj, 141, 30

\bibitem[{{Crossfield} {et~al.}(2010){Crossfield}, {Hansen}, {Harrington},
  {Cho}, {Deming}, {Menou}, \& {Seager}}]{Crossfield_2010}
{Crossfield}, I.~J.~M., {Hansen}, B.~M.~S., {Harrington}, J., {Cho}, J.,
  {Deming}, D., {Menou}, K., \& {Seager}, S. 2010, \apj, 723, 1436

\bibitem[Crossfield et al.(2012)]{Crossfield_2012} Crossfield, 
I.~J.~M., Hansen, B.~M.~S., \& Barman, T.\ 2012, arXiv:1201.1023 

\bibitem[{{Deming} {et~al.}(2011){Deming}, {Knutson}, {Agol}, {Desert},
  {Burrows}, {Fortney}, {Charbonneau}, {Cowan}, {Laughlin}, {Langton},
  {Showman}, \& {Lewis}}]{Deming_2011}
{Deming}, D., {Knutson}, H., {Agol}, E., {Desert}, J.-M., {Burrows}, A.,
  {Fortney}, J.~J., {Charbonneau}, D., {Cowan}, N.~B., {Laughlin}, G.,
  {Langton}, J., {Showman}, A.~P., \& {Lewis}, N.~K. 2011, \apj, 726, 95

\bibitem[{{Demory} {et~al.}(2011){Demory}, {Seager}, {Madhusudhan}, {Kjeldsen},
  {Christensen-Dalsgaard}, {Gillon}, {Rowe}, {Welsh}, {Adams}, {Dupree},
  {McCarthy}, {Kulesa}, {Borucki}, \& {Koch}}]{Demory_2011}
{Demory}, B.-O., {Seager}, S., {Madhusudhan}, N., {Kjeldsen}, H.,
  {Christensen-Dalsgaard}, J., {Gillon}, M., {Rowe}, J.~F., {Welsh}, W.~F.,
  {Adams}, E.~R., {Dupree}, A., {McCarthy}, D., {Kulesa}, C., {Borucki}, W.~J.,
  \& {Koch}, D.~G. 2011, \apjl, 735, L12+

\bibitem[{{D{\'e}sert} {et~al.}(2011{\natexlab{a}}){D{\'e}sert}, {Charbonneau},
  {Demory}, {Ballard}, {Carter}, {Fortney}, {Cochran}, {Endl}, {Quinn},
  {Isaacson}, {Fressin}, {Buchhave}, {Latham}, {Knutson}, {Bryson}, {Torres},
  {Rowe}, {Batalha}, {Borucki}, {Brown}, {Caldwell}, {Christiansen}, {Deming},
  {Fabrycky}, {Ford}, {Gilliland}, {Gillon}, {Haas}, {Jenkins}, {Kinemuchi},
  {Koch}, {Lissauer}, {Mullally}, {MacQueen}, {Marcy}, {Sasselov}, {Seager},
  {Still}, {Tenenbaum}, {Uddin}, \& {Winn}}]{Desert_2011b}
{D{\'e}sert}, J.-M., {Charbonneau}, D., {Demory}, B.-O., {Ballard}, S.,
  {Carter}, J.~A., {Fortney}, J.~J., {Cochran}, W.~D., {Endl}, M., {Quinn},
  S.~N., {Isaacson}, H.~T., {Fressin}, F., {Buchhave}, L.~A., {Latham}, D.~W.,
  {Knutson}, H.~A., {Bryson}, S.~T., {Torres}, G., {Rowe}, J.~F., {Batalha},
  N.~M., {Borucki}, W.~J., {Brown}, T.~M., {Caldwell}, D.~A., {Christiansen},
  J.~L., {Deming}, D., {Fabrycky}, D.~C., {Ford}, E.~B., {Gilliland}, R.~L.,
  {Gillon}, M., {Haas}, M.~R., {Jenkins}, J.~M., {Kinemuchi}, K., {Koch}, D.,
  {Lissauer}, J.~J., {Mullally}, F., {MacQueen}, P.~J., {Marcy}, G.~W.,
  {Sasselov}, D.~D., {Seager}, S., {Still}, M., {Tenenbaum}, P., {Uddin}, K.,
  \& {Winn}, J.~N. 2011{\natexlab{a}}, ArXiv e-prints

\bibitem[{{D{\'e}sert} {et~al.}(2011{\natexlab{b}}){D{\'e}sert}, {Sing},
  {Vidal-Madjar}, {H{\'e}brard}, {Ehrenreich}, {Lecavelier Des Etangs},
  {Parmentier}, {Ferlet}, \& {Henry}}]{Desert_2011}
{D{\'e}sert}, J.-M., {Sing}, D., {Vidal-Madjar}, A., {H{\'e}brard}, G.,
  {Ehrenreich}, D., {Lecavelier Des Etangs}, A., {Parmentier}, V., {Ferlet},
  R., \& {Henry}, G.~W. 2011{\natexlab{b}}, \aap, 526, A12+

\bibitem[{{Faigler} \& {Mazeh}(2011)}]{Faigler_2011}
{Faigler}, S. \& {Mazeh}, T. 2011, ArXiv e-prints

\bibitem[{{Fazio} {et~al.}(2004){Fazio}, {Hora}, {Allen}, {Ashby}, {Barmby},
  {Deutsch}, {Huang}, {Kleiner}, {Marengo}, {Megeath}, {Melnick}, {Pahre},
  {Patten}, {Polizotti}, {Smith}, {Taylor}, {Wang}, {Willner}, {Hoffmann},
  {Pipher}, {Forrest}, {McMurty}, {McCreight}, {McKelvey}, {McMurray}, {Koch},
  {Moseley}, {Arendt}, {Mentzell}, {Marx}, {Losch}, {Mayman}, {Eichhorn},
  {Krebs}, {Jhabvala}, {Gezari}, {Fixsen}, {Flores}, {Shakoorzadeh}, {Jungo},
  {Hakun}, {Workman}, {Karpati}, {Kichak}, {Whitley}, {Mann}, {Tollestrup},
  {Eisenhardt}, {Stern}, {Gorjian}, {Bhattacharya}, {Carey}, {Nelson},
  {Glaccum}, {Lacy}, {Lowrance}, {Laine}, {Reach}, {Stauffer}, {Surace},
  {Wilson}, {Wright}, {Hoffman}, {Domingo}, \& {Cohen}}]{Fazio_2004}
{Fazio}, G.~G., {Hora}, J.~L., {Allen}, L.~E., {Ashby}, M.~L.~N., {Barmby}, P.,
  {Deutsch}, L.~K., {Huang}, J.-S., {Kleiner}, S., {Marengo}, M., {Megeath},
  S.~T., {Melnick}, G.~J., {Pahre}, M.~A., {Patten}, B.~M., {Polizotti}, J.,
  {Smith}, H.~A., {Taylor}, R.~S., {Wang}, Z., {Willner}, S.~P., {Hoffmann},
  W.~F., {Pipher}, J.~L., {Forrest}, W.~J., {McMurty}, C.~W., {McCreight},
  C.~R., {McKelvey}, M.~E., {McMurray}, R.~E., {Koch}, D.~G., {Moseley}, S.~H.,
  {Arendt}, R.~G., {Mentzell}, J.~E., {Marx}, C.~T., {Losch}, P., {Mayman}, P.,
  {Eichhorn}, W., {Krebs}, D., {Jhabvala}, M., {Gezari}, D.~Y., {Fixsen},
  D.~J., {Flores}, J., {Shakoorzadeh}, K., {Jungo}, R., {Hakun}, C., {Workman},
  L., {Karpati}, G., {Kichak}, R., {Whitley}, R., {Mann}, S., {Tollestrup},
  E.~V., {Eisenhardt}, P., {Stern}, D., {Gorjian}, V., {Bhattacharya}, B.,
  {Carey}, S., {Nelson}, B.~O., {Glaccum}, W.~J., {Lacy}, M., {Lowrance},
  P.~J., {Laine}, S., {Reach}, W.~T., {Stauffer}, J.~A., {Surace}, J.~A.,
  {Wilson}, G., {Wright}, E.~L., {Hoffman}, A., {Domingo}, G., \& {Cohen}, M.
  2004, \apjs, 154, 10

\bibitem[{{Fortney} {et~al.}(2010){Fortney}, {Shabram}, {Showman}, {Lian},
  {Freedman}, {Marley}, \& {Lewis}}]{Fortney_2010}
{Fortney}, J.~J., {Shabram}, M., {Showman}, A.~P., {Lian}, Y., {Freedman},
  R.~S., {Marley}, M.~S., \& {Lewis}, N.~K. 2010, \apj, 709, 1396

\bibitem[{{Fossati} {et~al.}(2010){Fossati}, {Haswell}, {Froning}, {Hebb},
  {Holmes}, {Kolb}, {Helling}, {Carter}, {Wheatley}, {Collier Cameron},
  {Loeillet}, {Pollacco}, {Street}, {Stempels}, {Simpson}, {Udry}, {Joshi},
  {West}, {Skillen}, \& {Wilson}}]{Fossati_2010}
{Fossati}, L., {Haswell}, C.~A., {Froning}, C.~S., {Hebb}, L., {Holmes}, S.,
  {Kolb}, U., {Helling}, C., {Carter}, A., {Wheatley}, P., {Collier Cameron},
  A., {Loeillet}, B., {Pollacco}, D., {Street}, R., {Stempels}, H.~C.,
  {Simpson}, E., {Udry}, S., {Joshi}, Y.~C., {West}, R.~G., {Skillen}, I., \&
  {Wilson}, D. 2010, \apjl, 714, L222

\bibitem[{{Hanel} {et~al.}(1992){Hanel}, {Conrath}, {Jennings}, \&
  {Samuelson}}]{Hanel_1992}
{Hanel}, R.~A., {Conrath}, B.~J., {Jennings}, D.~E., \& {Samuelson}, R.~E.
  1992, in Exploration of the solar system by infrared remote sensing Cambridge
  University Press (Cambridge Planetary Science Series, No.~7), 484 p., ed.
  {Hanel, R.~A., Conrath, B.~J., Jennings, D.~E., \& Samuelson, R.~E. }

\bibitem[{{Hebb} {et~al.}(2009){Hebb}, {Collier-Cameron}, {Loeillet},
  {Pollacco}, {H{\'e}brard}, {Street}, {Bouchy}, {Stempels}, {Moutou},
  {Simpson}, {Udry}, {Joshi}, {West}, {Skillen}, {Wilson}, {McDonald},
  {Gibson}, {Aigrain}, {Anderson}, {Benn}, {Christian}, {Enoch}, {Haswell},
  {Hellier}, {Horne}, {Irwin}, {Lister}, {Maxted}, {Mayor}, {Norton}, {Parley},
  {Pont}, {Queloz}, {Smalley}, \& {Wheatley}}]{Hebb_2009}
{Hebb}, L., {Collier-Cameron}, A., {Loeillet}, B., {Pollacco}, D.,
  {H{\'e}brard}, G., {Street}, R.~A., {Bouchy}, F., {Stempels}, H.~C.,
  {Moutou}, C., {Simpson}, E., {Udry}, S., {Joshi}, Y.~C., {West}, R.~G.,
  {Skillen}, I., {Wilson}, D.~M., {McDonald}, I., {Gibson}, N.~P., {Aigrain},
  S., {Anderson}, D.~R., {Benn}, C.~R., {Christian}, D.~J., {Enoch}, B.,
  {Haswell}, C.~A., {Hellier}, C., {Horne}, K., {Irwin}, J., {Lister}, T.~A.,
  {Maxted}, P., {Mayor}, M., {Norton}, A.~J., {Parley}, N., {Pont}, F.,
  {Queloz}, D., {Smalley}, B., \& {Wheatley}, P.~J. 2009, \apj, 693, 1920

\bibitem[{{Husnoo} {et~al.}(2011){Husnoo}, {Pont}, {H{\'e}brard}, {Simpson},
  {Mazeh}, {Bouchy}, {Moutou}, {Arnold}, {Boisse}, {D{\'{\i}}az},
  {Eggenberger}, \& {Shporer}}]{Husnoo_2011}
{Husnoo}, N., {Pont}, F., {H{\'e}brard}, G., {Simpson}, E., {Mazeh}, T.,
  {Bouchy}, F., {Moutou}, C., {Arnold}, L., {Boisse}, I., {D{\'{\i}}az}, R.~F.,
  {Eggenberger}, A., \& {Shporer}, A. 2011, \mnras, 413, 2500

\bibitem[{{Iro} {et~al.}(2005){Iro}, {B{\'e}zard}, \& {Guillot}}]{Iro_2005}
{Iro}, N., {B{\'e}zard}, B., \& {Guillot}, T. 2005, \aap, 436, 719

\bibitem[{{Kipping} \& {Spiegel}(2011)}]{Kipping_2011}
{Kipping}, D.~M. \& {Spiegel}, D.~S. 2011, ArXiv e-prints

\bibitem[{{Knutson} {et~al.}(2007){Knutson}, {Charbonneau}, {Allen}, {Fortney},
  {Agol}, {Cowan}, {Showman}, {Cooper}, \& {Megeath}}]{Knutson_2007a}
{Knutson}, H.~A., {Charbonneau}, D., {Allen}, L.~E., {Fortney}, J.~J., {Agol},
  E., {Cowan}, N.~B., {Showman}, A.~P., {Cooper}, C.~S., \& {Megeath}, S.~T.
  2007, \nat, 447, 183

\bibitem[{{Knutson} {et~al.}(2009{\natexlab{a}}){Knutson}, {Charbonneau},
  {Burrows}, {O'Donovan}, \& {Mandushev}}]{Knutson_2009b}
{Knutson}, H.~A., {Charbonneau}, D., {Burrows}, A., {O'Donovan}, F.~T., \&
  {Mandushev}, G. 2009{\natexlab{a}}, \apj, 691, 866

\bibitem[{{Knutson} {et~al.}(2009{\natexlab{b}}){Knutson}, {Charbonneau},
  {Cowan}, {Fortney}, {Showman}, {Agol}, {Henry}, {Everett}, \&
  {Allen}}]{Knutson_2009a}
{Knutson}, H.~A., {Charbonneau}, D., {Cowan}, N.~B., {Fortney}, J.~J.,
  {Showman}, A.~P., {Agol}, E., {Henry}, G.~W., {Everett}, M.~E., \& {Allen},
  L.~E. 2009{\natexlab{b}}, \apj, 690, 822

\bibitem[{{Knutson} {et~al.}(2011){Knutson}, {Madhusudhan}, {Cowan},
  {Christiansen}, {Agol}, {Deming}, {D{\'e}sert}, {Charbonneau}, {Henry},
  {Homeier}, {Langton}, {Laughlin}, \& {Seager}}]{Knutson_2011}
{Knutson}, H.~A., {Madhusudhan}, N., {Cowan}, N.~B., {Christiansen}, J.~L.,
  {Agol}, E., {Deming}, D., {D{\'e}sert}, J.-M., {Charbonneau}, D., {Henry},
  G.~W., {Homeier}, D., {Langton}, J., {Laughlin}, G., \& {Seager}, S. 2011,
  \apj, 735, 27

\bibitem[{{Kopparapu} {et~al.}(2011){Kopparapu}, {Kasting}, \&
  {Zahnle}}]{Kopparapu_2011}
{Kopparapu}, R.~k., {Kasting}, J.~F., \& {Zahnle}, K.~J. 2011, ArXiv e-prints

\bibitem[{{Kurucz}(1979)}]{Kurucz_1979}
{Kurucz}, R.~L. 1979, \apjs, 40, 1

\bibitem[{{Kurucz}(2005)}]{Kurucz_2005}
---. 2005, Memorie della Societa Astronomica Italiana Supplement, 8, 14

\bibitem[{{Lai} {et~al.}(2010){Lai}, {Helling}, \& {van den Heuvel}}]{Lai_2010}
{Lai}, D., {Helling}, C., \& {van den Heuvel}, E.~P.~J. 2010, \apj, 721, 923

\bibitem[{{Leconte} {et~al.}(2011){Leconte}, {Lai}, \&
  {Chabrier}}]{Leconte_2011}
{Leconte}, J., {Lai}, D., \& {Chabrier}, G. 2011, \aap, 528, A41+

\bibitem[{{Li} {et~al.}(2010){Li}, {Miller}, {Lin}, \& {Fortney}}]{Li_2010}
{Li}, S.-L., {Miller}, N., {Lin}, D.~N.~C., \& {Fortney}, J.~J. 2010, \nat,
  463, 1054

\bibitem[{{Llama} {et~al.}(2011){Llama}, {Wood}, {Jardine}, {Vidotto},
  {Helling}, {Fossati}, \& {Haswell}}]{Llama_2011}
{Llama}, J., {Wood}, K., {Jardine}, M., {Vidotto}, A.~A., {Helling}, C.,
  {Fossati}, L., \& {Haswell}, C.~A. 2011, \mnras, L283+

\bibitem[{{Loeb} \& {Gaudi}(2003)}]{Loeb_2003}
{Loeb}, A. \& {Gaudi}, B.~S. 2003, \apjl, 588, L117

\bibitem[{{L{\'o}pez-Morales} {et~al.}(2010){L{\'o}pez-Morales}, {Coughlin},
  {Sing}, {Burrows}, {Apai}, {Rogers}, {Spiegel}, \&
  {Adams}}]{Lopez-Morales_2010}
{L{\'o}pez-Morales}, M., {Coughlin}, J.~L., {Sing}, D.~K., {Burrows}, A.,
  {Apai}, D., {Rogers}, J.~C., {Spiegel}, D.~S., \& {Adams}, E.~R. 2010, \apjl,
  716, L36

\bibitem[{{Maciejewski} {et~al.}(2011){Maciejewski}, {Errmann}, {Raetz},
  {Seeliger}, {Spaleniak}, \& {Neuh{\"a}user}}]{Maciejewski_2011}
{Maciejewski}, G., {Errmann}, R., {Raetz}, S., {Seeliger}, M., {Spaleniak}, I.,
  \& {Neuh{\"a}user}, R. 2011, \aap, 528, A65+

\bibitem[{{Madhusudhan} {et~al.}(2011){Madhusudhan}, {Harrington}, {Stevenson},
  {Nymeyer}, {Campo}, {Wheatley}, {Deming}, {Blecic}, {Hardy}, {Lust},
  {Anderson}, {Collier-Cameron}, {Britt}, {Bowman}, {Hebb}, {Hellier},
  {Maxted}, {Pollacco}, \& {West}}]{Madhusudhan_2011}
{Madhusudhan}, N., {Harrington}, J., {Stevenson}, K.~B., {Nymeyer}, S.,
  {Campo}, C.~J., {Wheatley}, P.~J., {Deming}, D., {Blecic}, J., {Hardy},
  R.~A., {Lust}, N.~B., {Anderson}, D.~R., {Collier-Cameron}, A., {Britt},
  C.~B.~T., {Bowman}, W.~C., {Hebb}, L., {Hellier}, C., {Maxted}, P.~F.~L.,
  {Pollacco}, D., \& {West}, R.~G. 2011, \nat, 469, 64

\bibitem[{{Madhusudhan} \& {Seager}(2009)}]{Madhusudhan_2009b}
{Madhusudhan}, N. \& {Seager}, S. 2009, \apj, 707, 24

\bibitem[{{Mandel} \& {Agol}(2002)}]{Mandel_2002}
{Mandel}, K. \& {Agol}, E. 2002, \apjl, 580, L171

\bibitem[{{Menou}(2011)}]{Menou_2011}
{Menou}, K. 2011, ArXiv e-prints

\bibitem[{{Morales-Calder{\'o}n} {et~al.}(2006){Morales-Calder{\'o}n},
  {Stauffer}, {Kirkpatrick}, {Carey}, {Gelino}, {Barrado y Navascu{\'e}s},
  {Rebull}, {Lowrance}, {Marley}, {Charbonneau}, {Patten}, {Megeath}, \&
  {Buzasi}}]{Morales_2006}
{Morales-Calder{\'o}n}, M., {Stauffer}, J.~R., {Kirkpatrick}, J.~D., {Carey},
  S., {Gelino}, C.~R., {Barrado y Navascu{\'e}s}, D., {Rebull}, L., {Lowrance},
  P., {Marley}, M.~S., {Charbonneau}, D., {Patten}, B.~M., {Megeath}, S.~T., \&
  {Buzasi}, D. 2006, \apj, 653, 1454

\bibitem[{{Perna} {et~al.}(2010){Perna}, {Menou}, \& {Rauscher}}]{Perna_2010}
{Perna}, R., {Menou}, K., \& {Rauscher}, E. 2010, \apj, 719, 1421

\bibitem[{{Pont} {et~al.}(2008){Pont}, {Knutson}, {Gilliland}, {Moutou}, \&
  {Charbonneau}}]{Pont_2008}
{Pont}, F., {Knutson}, H., {Gilliland}, R.~L., {Moutou}, C., \& {Charbonneau},
  D. 2008, \mnras, 385, 109

\bibitem[{{Pont} {et~al.}(2006){Pont}, {Zucker}, \& {Queloz}}]{Pont_2006}
{Pont}, F., {Zucker}, S., \& {Queloz}, D. 2006, \mnras, 373, 231

\bibitem[{{Ragozzine} \& {Wolf}(2009)}]{Ragozzine_2009}
{Ragozzine}, D. \& {Wolf}, A.~S. 2009, \apj, 698, 1778

\bibitem[{{Rauscher} {et~al.}(2007){Rauscher}, {Menou}, {Seager}, {Deming},
  {Cho}, \& {Hansen}}]{Rauscher_2007}
{Rauscher}, E., {Menou}, K., {Seager}, S., {Deming}, D., {Cho}, J.~Y.-K., \&
  {Hansen}, B.~M.~S. 2007, \apj, 664, 1199

\bibitem[{{Rowe} {et~al.}(2008){Rowe}, {Matthews}, {Seager}, {Miller-Ricci},
  {Sasselov}, {Kuschnig}, {Guenther}, {Moffat}, {Rucinski}, {Walker}, \&
  {Weiss}}]{Rowe_2008}
{Rowe}, J.~F., {Matthews}, J.~M., {Seager}, S., {Miller-Ricci}, E., {Sasselov},
  D., {Kuschnig}, R., {Guenther}, D.~B., {Moffat}, A.~F.~J., {Rucinski}, S.~M.,
  {Walker}, G.~A.~H., \& {Weiss}, W.~W. 2008, \apj, 689, 1345

\bibitem[{{Santerne} {et~al.}(2011){Santerne}, {Bonomo}, {H{\'e}brard},
  {Deleuil}, {Moutou}, {Almenara}, {Bouchy}, \& {D{\'{\i}}az}}]{Saterne_2011}
{Santerne}, A., {Bonomo}, A.~S., {H{\'e}brard}, G., {Deleuil}, M., {Moutou},
  C., {Almenara}, J.~., {Bouchy}, F., \& {D{\'{\i}}az}, R.~F. 2011, ArXiv
  e-prints

\bibitem[{Schwarz(1978)}]{Schwarz_1978}
Schwarz, G. 1978, The Annals of Statistics, 6, pp. 461

\bibitem[{{Seager} {et~al.}(2005){Seager}, {Richardson}, {Hansen}, {Menou},
  {Cho}, \& {Deming}}]{Seager_2005}
{Seager}, S., {Richardson}, L.~J., {Hansen}, B.~M.~S., {Menou}, K., {Cho},
  J.~Y.-K., \& {Deming}, D. 2005, \apj, 632, 1122

\bibitem[{{Sing} {et~al.}(2011){Sing}, {Pont}, {Aigrain}, {Charbonneau},
  {D{\'e}sert}, {Gibson}, {Gilliland}, {Hayek}, {Henry}, {Knutson}, {Lecavelier
  Des Etangs}, {Mazeh}, \& {Shporer}}]{Sing_2011}
{Sing}, D.~K., {Pont}, F., {Aigrain}, S., {Charbonneau}, D., {D{\'e}sert},
  J.-M., {Gibson}, N., {Gilliland}, R., {Hayek}, W., {Henry}, G., {Knutson},
  H., {Lecavelier Des Etangs}, A., {Mazeh}, T., \& {Shporer}, A. 2011, \mnras,
  1159

\bibitem[{{Stevenson} {et~al.}(2011){Stevenson}, {Harrington}, {Fortney},
  {Loredo}, {Hardy}, {Nymeyer}, {Bowman}, {Cubillos}, {Bowman}, \&
  {Hardin}}]{Stevenson_2011}
{Stevenson}, K.~B., {Harrington}, J., {Fortney}, J., {Loredo}, T.~J., {Hardy},
  R.~A., {Nymeyer}, S., {Bowman}, W.~C., {Cubillos}, P., {Bowman}, M.~O., \&
  {Hardin}, M. 2011, ArXiv e-prints

\bibitem[{Vickers(1996)}]{Vickers_1996}
Vickers, G.~T. 1996, Powder Technology, 86, 195

\bibitem[{{Vidotto} {et~al.}(2010){Vidotto}, {Jardine}, \&
  {Helling}}]{Vidotto_2010}
{Vidotto}, A.~A., {Jardine}, M., \& {Helling}, C. 2010, \apjl, 722, L168

\bibitem[{{von Zeipel}(1924)}]{VonZeipel_1924}
{von Zeipel}, H. 1924, \mnras, 84, 665

\bibitem[{{Watkins} \& {Cho}(2010)}]{Watkins_2010}
{Watkins}, C. \& {Cho}, J.~Y.-K. 2010, \apj, 714, 904

\bibitem[{{Welsh} {et~al.}(2010){Welsh}, {Orosz}, {Seager}, {Fortney},
  {Jenkins}, {Rowe}, {Koch}, \& {Borucki}}]{Welsh_2010}
{Welsh}, W.~F., {Orosz}, J.~A., {Seager}, S., {Fortney}, J.~J., {Jenkins}, J.,
  {Rowe}, J.~F., {Koch}, D., \& {Borucki}, W.~J. 2010, \apjl, 713, L145

\bibitem[{{Werner} {et~al.}(2004){Werner}, {Roellig}, {Low}, {Rieke}, {Rieke},
  {Hoffmann}, {Young}, {Houck}, {Brandl}, {Fazio}, {Hora}, {Gehrz}, {Helou},
  {Soifer}, {Stauffer}, {Keene}, {Eisenhardt}, {Gallagher}, {Gautier}, {Irace},
  {Lawrence}, {Simmons}, {Van Cleve}, {Jura}, {Wright}, \&
  {Cruikshank}}]{Werner_2004}
{Werner}, M.~W., {Roellig}, T.~L., {Low}, F.~J., {Rieke}, G.~H., {Rieke}, M.,
  {Hoffmann}, W.~F., {Young}, E., {Houck}, J.~R., {Brandl}, B., {Fazio}, G.~G.,
  {Hora}, J.~L., {Gehrz}, R.~D., {Helou}, G., {Soifer}, B.~T., {Stauffer}, J.,
  {Keene}, J., {Eisenhardt}, P., {Gallagher}, D., {Gautier}, T.~N., {Irace},
  W., {Lawrence}, C.~R., {Simmons}, L., {Van Cleve}, J.~E., {Jura}, M.,
  {Wright}, E.~L., \& {Cruikshank}, D.~P. 2004, \apjs, 154, 1

\bibitem[{{Williams} {et~al.}(2006){Williams}, {Charbonneau}, {Cooper},
  {Showman}, \& {Fortney}}]{Williams_2006}
{Williams}, P.~K.~G., {Charbonneau}, D., {Cooper}, C.~S., {Showman}, A.~P., \&
  {Fortney}, J.~J. 2006, \apj, 649, 1020

\bibitem[{{Winn}(2010)}]{Winn_2010}
{Winn}, J.~N. 2010, in Exoplanets, ed. {Seager, S.}, 55--77

\bibitem[{{Winn} {et~al.}(2008){Winn}, {Henry}, {Torres}, \&
  {Holman}}]{Winn_2008}
{Winn}, J.~N., {Henry}, G.~W., {Torres}, G., \& {Holman}, M.~J. 2008, \apj,
  675, 1531

\bibitem[{{Winn} {et~al.}(2007){Winn}, {Holman}, \& {Roussanova}}]{Winn_2007}
{Winn}, J.~N., {Holman}, M.~J., \& {Roussanova}, A. 2007, \apj, 657, 1098

\bibitem[{{Zhao} {et~al.}(2011){Zhao}, {Monnier}, {Swain}, {Barman}, \&
  {Hinkley}}]{Zhao_2011}
{Zhao}, M., {Monnier}, J.~D., {Swain}, M.~R., {Barman}, T., \& {Hinkley}, S.
  2011, ArXiv e-prints

\end{thebibliography}
\end{document}